\pdfoutput=1

\documentclass[12pt]{article}

\usepackage{amsfonts}
\usepackage{amsmath}
\usepackage{amssymb}
\usepackage{bigints}
\usepackage{booktabs}
\usepackage[nosort]{cite}
\usepackage{color}
\usepackage{dsfont}
\usepackage{float}
\usepackage{framed}
\usepackage{graphicx}
\usepackage{indentfirst}
\usepackage{mathrsfs}
\usepackage{multirow}
\usepackage{pdflscape}
\usepackage{setspace}
\usepackage{subdepth}
\usepackage{subfig}
\usepackage{titlesec}
\usepackage[dotinlabels]{titletoc}
\usepackage{wrapfig}
\usepackage[all]{xy}
\usepackage{young}
\usepackage[vcentermath]{youngtab}
\usepackage{relsize}
\usepackage{stackengine}

\usepackage{hyperref}

\numberwithin{equation}{section}

\usepackage{verbatim}

 \newcommand{\reef}[1]{(\ref{#1})}

\newcommand{\be}{\begin{equation}}
\newcommand{\ee}{\end{equation}}

\newcommand{\bml}{\begin{multline}}
\newcommand{\emll}{\end{multline}}

\newcommand{\nn}{\nonumber}

\def\({\left(} \def\){\right)}
\def\[{\left[} \def\]{\right]}

\def\Re{\text{Re}}
\def\Im{\text{Im}}

\def\hG{\hat G}

\def\sgn{\text{sgn}}

\def\al{\alpha}

\def\mO{\mathcal{O}}

\def\eps{\epsilon}

\def\g{\gamma}

\def\lam{\lambda}

\newcommand{\G}{\Gamma}

\def\d{\partial}

\newcommand{\la}{\langle}
\newcommand{\ra}{\rangle}

\newcommand{\bea}{\begin{eqnarray}}
\newcommand{\eea}{\end{eqnarray}}

\newcommand{\bi}{\begin{itemize}}
\newcommand{\ei}{\end{itemize}}

\usepackage[left=2cm,right=2cm,top=2cm,bottom=2cm]{geometry}
\titleformat{\section}{\large\bfseries}{\thesection.}{4pt}{}
\titlespacing{\section}{0pt}{22pt}{6pt}

\titleformat{\subsection}{\large\bfseries}{\thesubsection.}{4pt}{}
\titlespacing{\subsection}{0pt}{18pt}{6pt}

\titleformat{\subsubsection}{\normalfont\bfseries}{\thesubsubsection.}{4pt}{}
\titlespacing{\subsubsection}{0pt}{16pt}{6pt}

\def\ie{\begin{equation}\begin{aligned}}
\def\fe{\end{aligned}\end{equation}}

\def\hat{\widehat}
\def\h{\hat}
\def\bar{\overline}

\def\d{\partial}

\def\1{{\mathds 1}}
\def\Re{\mathop{\rm Re}}
\def\Im{\mathop{\rm Im}}

\def\o{\omega}

\def\da{a^{\dagger}}

\def\sd{\mathscr{D}}
\def\bsd{\bar{\mathscr{D}}}

\DeclareFontShape{OT1}{cmr}{mx}{n}%
    {<->cmr10}{}
\newcommand{\mytitlefont}{\fontseries{mx}\selectfont}
\DeclareMathAlphabet{\titlemath}{OT1}{cmr}{mx}{n}

\usepackage[UKenglish]{datetime}

\def\sec{\section}
\def\ss{\subsection}
\def\sss{\subsubsection}

\begin{document}

\begin{titlepage}

\begin{center}

~\\[1cm]

{\fontsize{20pt}{0pt} \mytitlefont Feynman rules for forced wave turbulence}\\[10pt]

~\\[0.2cm]

{\fontsize{14pt}{0pt}Vladimir Rosenhaus{\small $^{1}$} and Michael Smolkin{\small $^{2}$}}

~\\[0.1cm]

\it{$^1$ Initiative for the Theoretical Sciences}\\ \it{ The Graduate Center, CUNY}\\ \it{
 365 Fifth Ave, New York, NY 10016, USA}\\[.5cm]
 
 \it{$^2$ The Racah Institute of Physics}\\ \it{The Hebrew University of Jerusalem} \\ \it{
Jerusalem 91904, Israel}

~\\[0.6cm]

\end{center}

\noindent 
It has long been known that weakly nonlinear field theories can have a late-time stationary state that is not the thermal state, but a wave turbulent state with a far-from-equilibrium cascade of energy. We go beyond the existence of the wave turbulent state, studying fluctuations about the wave turbulent state. Specifically, we take a classical field theory with an arbitrary quartic interaction and add dissipation and Gaussian-random forcing. Employing the path integral relation  between stochastic classical field theories and quantum field theories, we give a prescription, in terms of  Feynman diagrams, for computing correlation functions in this system.  We explicitly compute the two-point and four-point functions of the field to next-to-leading order in the coupling. Through an appropriate choice of forcing and dissipation, these correspond to correlation functions in the wave turbulent state. In particular, we derive the kinetic equation to next-to-leading order.

\vfill

\noindent \href{mailto:vrosenhaus@gc.cuny.edu}{vrosenhaus@gc.cuny.edu}\\ \href{mailto:michael.smolkin@mail.huji.ac.il}{michael.smolkin@mail.huji.ac.il}\\[5pt]
\end{titlepage}


\tableofcontents
~\\

\section{Introduction}

Statistical physics is built around the study of the thermal state. This is for good reason:  at late times, generic closed systems with generic initial conditions are, for most purposes,  indistinguishable from the thermal state. Nevertheless, there is an increasingly large range of contexts in which one needs to study far-from-equilibrium systems,  ranging from the quark gluon plasma produced heavy ion collisions \cite{Schlichting:2019abc} to quenches in cold atom experiments \cite{Langen:2014saa, Kett, Polkovnikov:2010yn}. This is challenging:  the thermal state is now irrelevant, and there is no clear  ``universal'' state to replace it; it appears one must study every  far-from-equilibrium initial condition on a case-by-case basis.

The key is to look at a subsystem, consisting of a range of  modes of the underlying field. A nonlinear system couples  modes of different wavenumber, and there will be a flux of modes passing through the subsystem; the subsystem behaves like an open system. As an approximation,  we can replace our subsystem with a simple open system: one in which there is a perpetual flux passing through, maintained by external forcing and dissipation acting on an otherwise closed system.~\footnote{This  approximation is valid for certain far-from-equilibrium initial conditions and for momenta within some range, and for intermediate times that are long after initial transients have decayed but well before equilibrium has been reached. Of course, systems with forcing and dissipation are by themselves physically relevant, so the motivation in terms of an intermediate stage in the thermalization process is not necessary.} This is the system we will study.

Once a system is open, it is liberated from the requirement of late time thermalization, allowing for rich late time behavior. Beyond the thermal state, the next simplest possible late time state is a stationary, but non-equilibrium, state. 
 We can control the  state through the microscopic parameters of the Hamiltonian -- the dispersion relation and the nonlinear interaction -- as well as the external forcing and dissipation.  Remarkably, even  for a weakly interacting nonlinear system, there are cases in which one can find  a stationary, non-equilibrium state - the Kolmogorov-Zakharov state \cite{Zakharov}. This is wave turbulence, or weak turbulence \cite{Falkovich, Nazarenko}.~\footnote{Wave turbulence is distinctly different from the more familiar hydrodynamics turbulence, see Appendix.~\ref{appx:quartic}.} It has been shown to occur in an incredible range of  contexts,  from surface gravity waves to waves on vibrating elastic plates,  to waves in the quark-gluon plasma produced after a heavy ion collision. 

Much of the work on wave turbulence has focused on establishing the existence and properties of the turbulent state.
A broader question is how to repeat everything we know in  statistical mechanics, but  based on the wave turbulent state instead of the thermal state. Concretely, how to characterize fluctuations \textit{about} the turbulent state. This work is a step in that direction. Specifically, we take a classical nonlinear field theory, with an arbitrary dispersion relation and arbitrary quartic interaction. We add dissipation, as well as external forcing, where the forcing function is drawn from a Gaussian distribution.  We give a general prescription for computing correlation functions of the field. The basic tool that we use is that a classical field theory with stochastic forcing is a quantum field theory, which in turn can be solved perturbatively through Feynman diagrams.

Wave turbulence is both an old topic \cite{Falkovich} and one under active study \cite{Nazarenko}. Questions of current interest include: new physical contexts exhibiting classical wave turbulence,~\footnote{This includes: elastic plate wave turbulence \cite{DURING201742, During, Mordant1, Mordant2, Chibbaro16, PhysRevFluids.4.064804, 2021Ben}, turbulence in surface gravity waves \cite{Zakharov1, Zakharov2, Onorato3, 2022Zhang, 2021Cab}, in gravitational waves \cite{gravity, 2021Naz, 2021Axel, PhysRevLett.96.204501}, acoustic turbulence\cite{griffin2021energy}, planetary Rossby waves \cite{2021Naz2}, waves on bubbles \cite{2021Paleg}, emergent hydrodynamics \cite{filho2021emergent}, in a diatomic chain \cite{pezzi2021threewave}, in plasmas  \cite{2021Zhu, 2021drift}, optical waves \cite{2019Fusaro,2021Bau, Baudin}, and rotating waves \cite{reun2020evidence, Cortet}, and in FPUT chains \cite{Onorato2}.    } wave turbulence in quantum mechanics and the nonlinear Schr\"odinger equation \cite{quantumTurb,buckmaster2021onset, banks2021direct, zhu2021testing, shukla2021nonequilibrium, 2021Nazz, DURING20091524,  2020Mul}, wave turbulence in quantum field theory and related concepts of prethermalization \cite{Micha:2004bv, Berges:2013lsa, Berges15, Schlicting, Chatrchyan:2020cxs ,Berges:2015kfa,Berges:2013eia,Berges:2004ce,Schmied:2018mte, Erne:2018gmz, Prufer, Glidden:2020qmu}, mathematical properties of wave turbulence \cite{Choi_2005,Eyink_2012, 
 Newell1, Newell2, Faou, Collot, dymov2020zakharov, Eynik2016} including properties of the kinetic equation \cite{Soffer, walton2021numerical, aceves2021wave, deng2021derivation, ampatzoglou2021derivation} and models such as \cite{MMT,ZAKHAROV20041, PhysRevFluids.2.052603, hrabski2021properties} and \cite{Falk2021} . 

Turning to work most relevant to this paper: the importance of quantities beyond the mode occupation number was  recently  stressed  in \cite{FalkovichShavit}. The question of higher order corrections (in the nonlinear interaction) to the expectation value of the mode occupation number, was studied in \cite{Erofeev}, though the expressions are unwieldy.  Through ingenious, if slightly mysterious, use of conservation laws,  \cite{Polyakov, Gurarie, Gurarie95} found the next-to-leading order correction to the kinetic equation governing the mode occupation number. Our results will reproduce and generalize the results of \cite{Polyakov, Gurarie, Gurarie95}, using a method that is relatively mundane and straightforward. The connection we employ, between stochastic classical field theories and quantum field theories, is well-known and such path integral methods have appeared before in turbulence, e.g. \cite{Migdal,  WYLD1961143, ZinnJustin, PhysRevE.54.4896, MSR}. However, as far we know, path integral methods have not been applied to classical wave turbulence for the explicit purpose of systematically computing correlation functions perturbatively in the coupling.~\footnote{There has been recent work on multimode statistics in the context of the random phase approximation  \cite{Choi_2005, Eyink_2012, Newell1, Newell2,  Nazarenko04, chibbaro20174wave, rosenzweig2021uniqueness, Jakobsen_2004}. }  We believe study along these lines will give a rich set of applications.

\sss*{Outline}

 In Sec.~\ref{sec2} we take a nonlinear wave equation with an added random forcing function. The most direct way of computing correlation functions is to take  a definite forcing function, solve the equations of motion for the field in terms of the forcing function, and then compute correlation function of the field in terms of correlation functions of the forcing function. In Sec.~\ref{sec:32} we formalize this procedure, which involves  a path integral over the field and a delta functional enforcing the equations of motion. The delta functional can itself be represented through a path integral over an auxiliary field. After integrating out the forcing function and the auxiliary field, one is left with simply a path integral for the field. This procedure shows that a classical field theory with Gaussian random forcing is equivalent to a quantum field theory, with a Lagrangian that is the square of the force-free equations of motion. The problem of computing correlation functions in stochastic field theory has transformed into a standard problem of computing correlation functions in a quantum field theory. 

In Sec.~\ref{sec4} we use this quantum field theory as our starting point. In standard quantum field theory, one computes vacuum correlation functions, where the vacuum is achieved by a small amount of evolution in Euclidean time (the $i\eps$ prescription). In our case, the correlation functions will automatically be computed in the stationary state of our choosing, maintained through forcing and dissipation. Indeed, the $i\eps$ is naturally present from the dissipation, and it does not have to be small. Due to dissipation, at late times the initial conditions become irrelevant and we gain time translation invariance. It is therefore beneficial to work in frequency space. In Sec.~\ref{sec:41} we use the Lagrangian to work out the Feynman rules for the propagator and the quartic and sextic interaction terms. In Sec.~\ref{sec:42} we compute the tree-level four-point function.

In Sec.~\ref{sec5} we compute one-loop diagrams. In Sec.~\ref{sec51} we compute the one-loop correction to the propagator, showing that it corresponds to a frequency shift. In Sec.~\ref{sec52} we compute the one-loop correction to the four-point function. In Sec.~\ref{sec53} we give an immediate application of these results: the kinetic equation to next-to-leading order, which encodes the evolution of the mode occupation number. The kinetic equation is the wave analog of the Boltzmann equation for particles in statistical mechanics. Obtaining next-to-leading order corrections to the Boltzmann equation is challenging, while here for the kinetic equation it is straightforward.

We conclude in Sec.~\ref{sec7} with a summary and ideas for future work. 

In Appendix~\ref{appx:quartic} we review wave turbulence, in particular the traditional derivation of the leading order kinetic equation, and the Kolmogorov-Zakharov turbulent cascade. The other appendices contain technical results relevant to the main body:  Appendix~\ref{ap:prop} derives several propagator identities,  Appendix~\ref{sec:43} derives the tree-level six-point function, Appendix~\ref{ap:integral} contains integrals used in the computation of the one-loop four-point function,  and Appendix~\ref{apE} is relevant to the next-to-leading order kinetic equation. 

\section{A nonlinear interacting field with random forcing} \label{sec2}
Our starting point is a  nonlinear classical field theory with a quartic interaction. It is best to work in Fourier space, with modes  $\phi_{k}$. It is common to do a canonical transformation, replacing the real field and momentum variables, $\phi_k$ and $\pi_k$, with the single complex variable $a_k$, whose complex conjugate we denote by $\da_k$, $\phi_k = \frac{1}{\sqrt{2 \o_k}}(a_k {+} \da_{k})$ and  $\pi_k = i\frac{\sqrt{\o_k}}{\sqrt{2}}(\da_k {-} a_{k})$. These variables are reminiscent of creation and annihilation operators in quantum mechanics, but we are of course just doing classical mechanics. The equations of motion are then first order, 
\be \label{eom}
\dot a_k + i\o_k  a_k =-2  i \sum_{p_2, p_3, p_4} \lam_{k p_2 p_3 p_4} \da_{p_2} a_{p_3} a_{p_4}~.
\ee
We may write the equations of motion in terms of a Hamiltonian, $
\dot{a}_k = -i \frac{\d H}{\d \da_k}  
$ where, 
\be \label{H21}
H = \sum_p \o_p \da_p a_p + \sum_{p_1, p_2, p_3, p_4}\!\! \lam_{p_1 p_2 p_3 p_4} \da_{p_1} \da_{p_2} a_{p_3} a_{p_4}~.
\ee
The symmetries of the coupling are $\lam_{p_1 p_2 p_3 p_4} = \lam_{p_2 p_1 p_3 p_4} =\lam_{p_1 p_2 p_4 p_3}  = \lam_{p_3 p_4 p_1 p_2}^*$.  Rather than writing an explicit momentum conserving delta function, we will just keep in mind that $\lam_{p_1 p_2 p_3 p_4} $ is only nonzero if $p_1{+} p_2 = p_3{+} p_4$. 

We  want to  add forcing $f_k(t)$  and dissipation $\gamma_k$ for mode $k$. The equations of motion become  \cite{Falkovichsummary}, 
\be   \label{eom1}
 \dot a_k = -i \frac{\d H}{\d \da_k}+ f_k(t) - \gamma_k a_k~.
 \ee
 At this stage the forcing term $f_k$ and the dissipation term $\gamma_k$ are arbitrary. We will want to average over the forcing. We take the forcing to be drawn from a Gaussian distribution,
 \be \label{Pf2}
 P\[f\] \sim \exp\( - \int d t \sum_k \frac{|f_k(t)|^2}{F_k}\)~, \ \ \ \  \la f_k(t) f_{p}^*(t')\ra = F_k \delta(k-p) \delta(t - t')~.
 \ee
We will work with arbitrary $F_k$ and $\gamma_k$, though in many contexts, one takes the forcing to be  nonvanishing only at low $k$ and the dissipation nonvanishing at high $k$, so that the forcing and dissipation don't directly affect the equations of motion for the modes in the inertial region in which there is a turbulent cascade.~\footnote{From the form of the probability distribution for $f_k$, in order to turn off $f_k$ for some mode $k$, one should take $F_k\rightarrow 0$, in which case $P\[f_k\] \rightarrow \delta(f_k)$.}

\sss*{Old-fashioned ``Diagrammatic'' method} \label{31}
The most straightforward way to solve the theory is through the so-called ``diagrammatic'' approach \cite{ZakharovLvov}: one solves the equations of motion with some definite forcing, and then averages over the forcing. While the path integral method we will  adopt in the next section will be computationally superior,  this straightforward method is conceptually clearer and serves as a useful check, so we briefly review it.  \\[-5pt]

\noindent\textit{Free theory:} Let us start with the free theory. The equations of motion  (\ref{eom1}) reduce to,
 \be \label{eq:free}
\dot a_k + ( i\omega_k + \gamma_k) a_k - f_k(t)= 0~,
\ee
and have the solution, 
\be \label{921}
a_k(t) =  e^{- (i\omega_k + \gamma_k) t}\( a_k(0) +  \int_0^t \!d t_1\, f_k(t_1) e^{(i \omega_k +\gamma_k)t_1}\)~.
\ee
We can use $a_k(t)$ with definite $f_k(t)$ to  compute correlation functions of $a_k(t)$ in terms of correlation functions of $f_k(t)$. For example, the two-point function is, 
\be
\!\!\!  \la \da_k(t_1) a_k(t_2)\ra  = e^{ i\omega_k t_{12} - \gamma_k(t_1+t_2) }\(  \la \da_k(0) a_k(0)\ra
  + \int_0^{t_1} \!d t'_1\int_0^{t_2}\! d t'_2\, \la f_k^*(t'_1) f_k(t_2')\ra \, e^{-i \o_k t_{12}' + \gamma_k(t_1' + t_2')} \)~,
\ee
where we used that $\la f\ra = 0$ and defined $t_{12} \equiv t_1 - t_2$. The occupation number of mode $k$ is denoted by $n_k(t)$.  Using $
\la f_k(t_1) f_k^*(t_2) \ra = F_k\, \delta(t_1 {-} t_2)$ gives,
\be \label{35}
  \la \da_k(t_1) a_k(t_2)\ra  = e^{i\omega_k t_{12} - \gamma_k(t_1+t_2) }\(  \la \da_k(0) a_k(0)\ra
  + \frac{ F_k }{2\gamma_k}\Big(\theta(t_{12}) e^{2\gamma_k t_2} + \theta(t_{21}) e^{2\gamma_k t_1} - 1 \Big) \)~,
\ee
where $\theta(t)$ is one for $t\geq 0$ and zero for $t<0$. 
If we take the late time limit: large $t_1$ and $t_2$ with finite $t_{12}$,  then the initial conditions $n_k(0)$ become irrelevant -- as one expects should be the case for a driven harmonic oscillator with damping --  and we get, 
\be \label{37}
  \la \da_k(t_1) a_k(t_2)\ra \rightarrow  \frac{ F_k }{2\gamma_k} e^{i\omega_k t_{12}  }
  \Big(\theta(t_{12}) e^{-\gamma_k t_{12}} + \theta(t_{21}) e^{\gamma_k t_{12}}  \Big)  =  \frac{ F_k }{2\gamma_k} e^{i\omega_k t_{12} - \gamma_k |t_{12}| } \ \ \ \text{as }  \  \ t\rightarrow \infty~.
\ee
Taking $t_2 = t_1$ in (\ref{35}) we get that $\la n_k(t)\ra$ is,
\be\label{nk}
\la n_k(t)\ra=  \la \da_k(0) a_k(0)\ra e^{- 2\gamma_k t} +\frac{ F_k }{2\gamma_k} (1 - e^{- 2\gamma_k t})  \rightarrow \frac{ F_k }{2\gamma_k} \equiv n_k~ \ \ \text{as }  \ \ t\rightarrow \infty~.
\ee
In what follows we will use $n_k$ as a shorthand for $F_k/2\g_k$. By appropriately picking $F_k$  and $\g_k$, one can achieve any $n_k$ that one desires. Likewise, one can take both $F_k$ and $\g_k$ to zero, while maintaining a finite ratio. \\[-5pt]

\noindent \textit{Interacting theory:}  If we restore interactions, we can solve the equations of motion perturbatively in the strength of the coupling. It is best to work in frequency space, $a_k(t) = \int \frac{d\omega}{2\pi} e^{- i\omega t} a_{k, \omega}$ and $ f_k(t) = \int \frac{d\omega}{2\pi} e^{- i\omega t} f_{k, \omega}$. 
The equations of motion (\ref{eom1}) become, 
\be
a_{k, \omega}
+2 i G_{k, \omega}\sum_{p_i, \o_i}   \delta_{\omega + \omega_1 - \omega_2 - \o_3} \lambda_{k p_1 p_2 p_3} \da_{p_1, \omega_1} a_{p_2, \omega_2}a_{p_3, \omega_3} =G_{k, \omega} f_{k,\omega}~, \ \ \ \ \ G_{k, \omega}  = \frac{i}{\omega- \omega_k + i \gamma_k}~.
\ee
Expanding $a_{k, \omega}$ perturbatively, 
$
a_{k,\omega} = a_{k, \omega}^{(0)} + a_{k, \omega}^{(1)} + \ldots~, 
$
the first two orders are given by, 
\bea \nn
a_{k,\o}^{(0)} &=& G_{k, \omega} f_{k,\omega}\\ 
a_{k, \omega}^{(1)} &=&-2 i G_{k, \omega}\!\!\! \sum_{p_1, p_2, p_3} \sum_{ \omega_1, \omega_2, \o_3}   \delta_{\omega + \omega_1 - \omega_2 - \o_3} \lambda_{k p_1 p_2 p_3} a^{\dagger\, (0)}_{p_1, \omega_1} a^{(0)}_{p_2, \omega_2}a^{(0)}_{p_3, \omega_3}~. 
\eea
Now, using the correlation function for the forcing, $
\la f_{k,\omega_1} f_{p, \omega_2}^*\ra  = F_k \delta_{k-p} 2 \pi \delta(\omega_1 {-} \omega_2), 
$
one may perturbatively compute correlation functions of $a_{k, \o}$.

This approach is straightforward, but tedious  \cite{Erofeev}. We would like to streamline the procedure, by integrating out the forcing at the outset. This is what we do next.

\ss{Path integral} \label{sec:32}

We are interested in computing correlation functions of products of various $a_k(t)$, such as e.g. the expectation value of the number operator, or multiple $a_k(t)$ with different momenta $k$ and inserted at different times $t$. 
The  expectation value of a general operator $\mO(a)$ is found by solving the equations of motion and computing $\mO(a)$ for each value of $f_k$, and then averaging over the $f_k$ as prescribed by the probability distribution $P\[f\]$ (\ref{Pf2}), 
\be
 \la \mO(a)\ra  = \int  \mathcal D f \mathcal D f^*\,  P\[f\] \mO(a)~, 
 \ee
 where one has to keep in mind that the $a_k(t)$ need to satisfy the equations of motion with the corresponding $f_k(t)$. 
Formally,  in order to ensure that we are using  $a_k(t)$ which satisfies the equations of motion for the chosen $f_k$, we introduce an integral over $a_k(t)$ and a delta function which ensures that the equations of motion are satisfied, see e.g. \cite{Migdal} or  \cite{WYLD1961143, ZinnJustin, PhysRevE.54.4896, MSR},
\bea \label{26}
  \la \mO(a)\ra &=& \int  \mathcal D a \mathcal D \da\, \mathcal D f \mathcal D f^* \, |J(a,\da)| \, P\[f\] \mO(a)\, \delta\big(\Re(E_f)\big)~ \delta\big(\Im(E_f)\big))~, 
  \\
  \text{E}_f &=& \dot a_k + i \frac{\delta H}{\delta \da_k}- f_k(t) +\gamma_k a_k ~,
  \nonumber
  \eea
where we introduced two delta functionals because $E_f$ is complex ($E_f = 0$ are the equations of motion),  and have let $|J(a,\da)|$ denote the modulus of the determinant of  the Jacobian,
\be
J(a,\da) = {\partial(\Re(E_f), \Im(E_f)) \over \partial (a, \da)}~.
\ee

Let us simplify \reef{26}. We start by noting that the Jacobian is actually one, $|J(a,\da)|=1$. To see this, we discretize time, sending $a_k(t)\to a_k^i$, and similarly for the other functions, such as the equation of motion operator $E_f(t)$, 
\be
\text{E}_f^i \Delta t= a_k^i - a_k^{i-1} + \Delta t \( i \frac{\delta H}{\delta a_k^{i-1}}- f_k^{i-1} +\gamma_k^{i-1} a_k^{i-1}\)~.
\ee
In this discretization, $J(a,\da)$  is a triangular matrix with unit diagonal, so the functional determinant equals unity, as claimed. 
Next we write the delta functionals in integral form as, 
  \be
\delta\big(\Re(E_f)\big)~ \delta\big(\Im(E_f)\big))  = \int \mathcal D \eta \mathcal D \eta^* \,  e^{i \int d t \sum_k \(\eta_k E_f^* + \eta_k^* E_f\)}~,
\ee 
so that $\la \mO(a)\ra$ becomes,
 \be
\la \mO(a)\ra = \!\! \int \mathcal D a \mathcal D \da\, \mathcal D f \mathcal D f^* \mathcal D \eta \mathcal D \eta^* ~ P\[f\] \mO(a) \exp\( i \int d t \sum_k \eta_k(t) E_f^*(t)+ \text{c.\,c.}\)~.
 \ee
Performing the integral over $f$ using $P\[f\]$ in (\ref{Pf2}) yields,
\be
 \la \mO(a)\ra = \int \mathcal D \eta \mathcal D \eta^* \mathcal D a \mathcal D \da\,\mO(a) \,  e^{-\int d t L}~,\ \ \ \ \ L = \sum_k \eta_k(t) F_k \eta_{k}^*(t) - i \eta_k(t) E_{f=0}^*
 - i \eta_k^*(t)E_{f=0}~,
\ee
where $\text{E}_{f=0}$ is the equation of motion term in (\ref{26}) without the forcing, $\text{E}_{f=0} = \dot a_k + i \frac{\delta H}{\delta a_k^*} +\gamma_k a_k $. 
Finally,  carrying out the Gaussian integral over $\eta_k$ gives,~\footnote{The integral over time in (\ref{Leff}) can be taken to extend to minus infinity in the past, but need only extend to the time at which $\mO(a)$ sits in the future: the integral arose from enforcing the equations of motion, however they do not need to be enforced at times later than where the observable is. }  
\be \label{Leff}
\la \mO(a)\ra = \int  \mathcal D a \mathcal D \da\, \mO(a) \, e^{-\int dt L}~, \ \ \ \ \ L = \sum_k \frac{|E_{f=0}|^2 }{F_k} ~.
\ee

The result is sensible: We started with equations of motion $E_{f=0}$,  added a forcing term $f_k$, and then averaged over a Gaussian distribution for the $f_k$. The result is an effective Lagrangian which is proportional to the magnitude squared of $E_{f=0}$. Notice that initially the only averaging was over the forcing term $f_k$, with $f_k$ having a Gaussian probability distribution. The end result, however, is the Lagrangian (\ref{Leff}) in which there is no forcing term, but with $a_k$  as a fluctuating variable; in other words, a quantum field theory. 

\sss*{The propagator}
The Gaussian theory is the part of the Lagrangian without the interaction term, 
 \bea \nn
 L_{free} = \sum_k \frac{1}{F_k} \Big|\sd_k a_k \Big|^2~, \ \ \ \ \ \ \ 
\sd_k a_k  &\equiv& \dot a_k + (i\o_k + \gamma_k) a_k~,\\ 
 \ \ \  \bsd_k a_k  &\equiv& \sd_k^* a_k= \dot a_k + (-i\o_k + \gamma_k) a_k~.  \label{Lfree}
\eea
The two-point function is given by  the inverse of the quadratic term,
 \be \label{prop1}
  \langle  \da_k(t_1)  a_p(t_2) \rangle =\delta_{k, p} F_k \[ \(-{d\over dt} + (i\omega_k + \gamma_k) \)
  \({d\over dt} + (-i\omega_k + \gamma_k) \) \]^{-1} ~,
 \ee 
which, through a Fourier transform,  gives,
\bea \label{324}
  \langle  \da_k(t_1)  a_p(t_2) \rangle &=& \delta_{k, p} F_k \, \int {d\omega \over 2\pi} {e^{i\omega t_{12}} \over 
   \big(-i\omega + (i\omega_k + \gamma_k) \big)  \big(i\omega + (-i\omega_k + \gamma_k) \big)} 
    \\
   &=&  { \delta_{k, p}  F_k\over 2\gamma_k } \[\theta(t_{12}) e^{i\omega_+ t_{12}}  + \theta(t_{21}) e^{i\omega_- t_{12}} 
  \] ~, 
   \quad \omega_\pm=\omega_k\pm i\gamma_k ~,\ \ \ \  t_{12} \equiv t_1 {-} t_2~.
   \nonumber
\eea
We may write this compactly as, 
 \be \label{prop}
D_k(t_{12})\equiv  \la \da_k(t_1) a_k(t_2)\ra  =n_k e^{(i \omega_k - \gamma_k \sgn(t_{12}))t_{12}}~,
  \ee
 where we used the definition (\ref{nk}) of $n_k$. Later on it will be convenient to use the shorthand $n_i \equiv n_{p_i}$. 
 
 The answer matches \textit{the late time limit} (\ref{37}) of  what we  found  (\ref{35}) through direct solution of the equations of motion. Had we wished, we could have precisely reproduced (\ref{35}), by accounting for  initial conditions when inverting $\sd_k$ in $L_{free}$. 
  The late time limit is important: it ensures that the propagator (\ref{prop}) only depends on the difference in times, and that Wick's theorem can be applied for finding higher-point correlation functions in the limit of vanishing coupling.

\sss*{Summary}
We have a nonlinear system with equations of motion (\ref{eom}) for the field $a_k(t)$. We add a forcing term to the equations of motion, and seek to average over the forcing drawn from a Gaussian-random distribution. A straightforward way is to solve the equations of motion with fixed forcing, and then average over the forcing. A formalization of this leads to a more efficient way, which we discussed in Sec.~\ref{sec:32}. The end result is that this problem is equivalent to the problem without forcing, where $a_k$ is a quantum field and the effective Lagrangian (\ref{Leff}) is the square of the original equations of motion without forcing.  This is what we will work with in the next section.

\section{Feynman rules and tree-level diagrams} \label{sec4}
In the previous section we found that the problem of a classical field theory with random forcing is equivalent to a quantum field theory, with a Lagrangian (\ref{Leff}). In this section we work out the Feynman rules and tree-level correlation functions. The Lagrangian and the Feynman rules are given in Sec.~\ref{sec:41}, and the tree-level four-point function is found in Sec.~\ref{sec:42}.

\ss{The  Lagrangian} \label{sec:41}
For the quartic theory with Hamiltonian (\ref{H21}) the equations of motion (\ref{eom1}) are $E_{f=0}=0$ where, 
\be
E_{f=0} = \sd_k a_k + 2  i \sum_{p_2, p_3, p_4} \lam_{k p_2 p_3 p_4} \da_{p_2} a_{p_3} a_{p_4}~,
\ee
where we used the definition of $\sd_k a_k$ given in (\ref{Lfree}). 
The Lagrangian (\ref{Leff}) we will be working with is therefore, 
\be \label{L42}
L = \sum_k \frac{|E_{f=0}|^2}{F_k} = \sum_k  \frac{1}{F_k} \Big| \sd_k a_k + 2  i \sum_{p_2, p_3, p_4} \lam_{k p_2 p_3 p_4} \da_{p_2} a_{p_3} a_{p_4}\Big|^2~.
\ee
 Let us expand this out, grouping terms by their power of the coupling. 
We have, 
\be \label{L33}
L = L_{free} + L_{\mO(\lam)} + L_{\mO(\lam^2)}~.
\ee
Here $L_{free}$ was given in (\ref{Lfree}). The term of order $\lam$ is,
\be
L_{\mO(\lam)} = 2 i \sum_{p_1, \ldots, p_4} \lam_{p_1\ldots p_4}  \frac{\bsd_{p_1}}{F_{p_1}}\da_{p_1} \da_{p_2} a_{p_3} a_{p_4} + \text{c.c.}~,
\ee
which we may write  as, 
\be
L_{\mO(\lam)} = 2 i \sum_{p_1, \ldots, p_4} \lam_{p_1\ldots p_4} \frac{\bsd_{p_1}}{F_{p_1}}\da_{p_1} \da_{p_2} a_{p_3} a_{p_4} - 2 i \sum_{p_1, \ldots, p_4} \lam_{p_1\ldots p_4} \frac{\sd_{p_3}}{F_{p_3}}\da_{p_1} \da_{p_2} a_{p_3} a_{p_4}~,
\ee
where for the last term we changed variables $p_1 \leftrightarrow p_3$, $p_2\leftrightarrow p_4$ and used $\lam_{p_1 p_2 p_3 p_4} =  \lam_{p_3 p_4 p_1 p_2}^*$. We may write this in a symmetric way, 
\be \label{quarticInt}
L_{\mO(\lam)} = i\sum_{p_1, \ldots, p_4} \lam_{p_1\ldots p_4}  \(  \frac{\bsd_{p_1}}{F_{p_1}}+  \frac{\bsd_{p_2}}{F_{p_2}} - \frac{\sd_{p_3}}{F_{p_3}} - \frac{\sd_{p_4}}{F_{p_4}}  \)\da_{p_1} \da_{p_2} a_{p_3} a_{p_4}~,
\ee
where one should remember that each $\sd_{p_i}$ only acts on the corresponding $a_{p_i}$. 
Finally, the $L_{\mO(\lam^2)}$ term is, 
\be \label{sexticInt}
L_{\mO(\lam^2)}  =4 \sum_{p_1, \ldots, p_7} \frac{1}{F_{p_7}} \lam_{p_1p_2p_3p_7} \lam_{p_7p_4p_5p_6} \da_{p_1}\da_{p_2} a_{p_3} \da_{p_4} a_{p_5} a_{p_6}~.
\ee
There isn't really a sum over $p_7$ here: recall that, in order to not write explicit momentum conserving delta functions, we have defined the couplings $\lam_{p_1 p_2 p_3 p_4}$ to be nonzero only when $p_1 {+}p_2 = p_{3}{+}p_4$. So in (\ref{sexticInt}) $p_7$ is fixed to be $p_7 = p_1{+}p_2{-} p_3 = p_5 {+} p_6 {-} p_4$. 

In total, the theory consists of a quadratic term (\ref{Lfree}), a quartic interaction (\ref{quarticInt}), and a sextic interaction (\ref{sexticInt}).

This looks just like any other quantum field theory. One distinction is the dissipation, $\g_k$ in $\sd_k a_k$. In fact, this is similar to the $i\eps$ in quantum field theory, see e.g. \cite{Peskin}: recall that in standard Lorentzian quantum field theory one wants to compute correlation functions in the vacuum. To achieve the vacuum (the vacuum of the full, interacting theory) one  evolves in a slightly imaginary time direction, so that the evolution operator $e^{- i H t}$ suppresses all states except the ground state. In frequency space, this corresponds to adding an $i\eps$ to the propagator. In our case we can keep the $\g_k$ finite, or take it to zero at the end, as is done with the $i\eps$. 

Let us work out the Feynman rules.

\sss*{Feynman rules}
The propagator $D_k(t_{12})\equiv  \la \da_k(t_1) a_k(t_2)\ra $ was given earlier, in (\ref{prop}), 
\be \label{prop38}
D_k(t_{12}) = n_k e^{i \omega_k t_{12} - \gamma_k |t_{12}|}~.
\ee
As is usually the case when there is time translation invariance, it will be easier to work in frequency space. The Fourier transform of the propagator is,
\be \label{DG}
D_{k, \o} \equiv \int dt D_{k}(t) e^{- i \o t}= F_k |G_{k, \o}|^2~, \ \ \ \ \   \text{where} \ \ \ \ \ \ G_{k, \o} = \frac{i}{\o{-} \o_{k} + i \gamma_{k}}~.
\ee

\begin{figure}
\centering
\subfloat[]{
\includegraphics[width=1in]{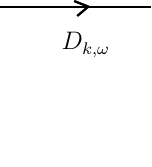}
}  \ \ \ \ \  \ \ \ \  \ \ \ \ 
\subfloat[]{
\includegraphics[width=1.2in]{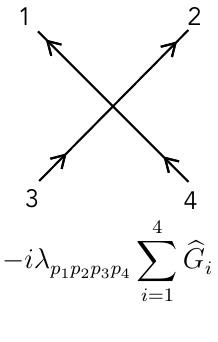}
} \ \ \ \ \ \ \ \ \ \ \ \ 
\subfloat[]{
\includegraphics[width=1.9in]{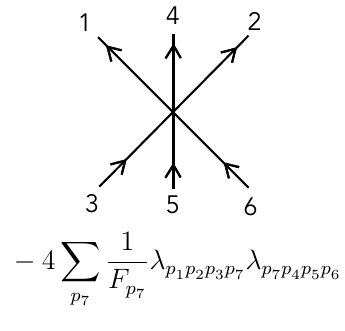}
}
\caption{Feynman rules corresponding to the Lagrangian (\ref{L33}). (a) The propagator (\ref{prop38}). (b) The quartic vertex (\ref{quarticV2}). (c) The sextic vertex (\ref{sext}). }  \label{FeyRules}
\end{figure}

The Feynman rules for the Lagrangian, in frequency space, are shown in Fig.~\ref{FeyRules}.
In particular, 
the quartic vertex  comes with a factor,~\footnote{Our Feynman rules are without the symmetrization factor. If one were to include the symmetrization factor in the Feynman rules, this would give an extra factor of $4$ in (\ref{quarticV2}) and an extra factor of $36$ in (\ref{sext}).}
\bml \label{quarticV2}
- i  \lam_{p_1 p_2 p_3 p_4}\! \!\[  \frac{(G_{p_1, \o_1}^{-1})^*}{F_{p_1}}{+}  \frac{(G_{p_2, \o_2}^{-1})^*}{F_{p_2}} {-} \frac{(G_{p_3, \o_3}^{-1})}{F_{p_3}} {-} \frac{(G_{p_4, \o_4}^{-1})}{F_{p_4}}  \] \, 2\pi  \delta(\o_1{+}\o_2{-}\o_3{-}\o_4)\\
 =- i \lam_{p_1 p_2 p_3 p_4}\sum_{i=1}^4 \hG_i\, \, 2\pi  \delta(\o_{1,2; 3,4})\,~, 
\end{multline}
\\[-20pt]
where, to simplify the notation,   we defined $\o_{i, j; k, l} \equiv \o_i {+} \o_j {-} \o_k{ -} \o_l$ and,
\be \label{Ghat}
\hG_i = \frac{1}{G^*_{p_i, \o_i} F_{p_i}}~, \ \ \ i = 1,2~, \ \ \ \ \ \ \ \hG_i = \frac{-1}{G_{p_i, \o_i} F_{p_i}}~, \ \ \ i =3, 4~.
\ee
The sextic vertex comes with a factor, 
\be \label{sext}
- 4 \sum_{p_7} \frac{1}{F_{p_7}} \lam_{p_1p_2p_3p_7} \lam_{p_7p_4p_5p_6}\  2\pi \delta(\o_{1,2,4;3,5,6})\,~,
\ee
where $\o_{i, j,k; l,m,n} \equiv \o_i{+}\o_j{+}\o_{k}{-}\o_l{-}\o_m{-}\o_n$ and, as we said earlier, the only term that contributes to the sum over $p_7$ is  $p_7= p_1+ p_2 - p_3$. 

We now start applying the Feynman rules to compute correlation functions. 

\ss{Four-point function} \label{sec:42}

The tree-level four-point function follows immediately from application of the Feynman rules. In frequency space it is given by,
\be\label{4ptF}
\!\!\!\la a_{p_1, \o_1} a_{p_2, \o_2} \da_{p_3, \o_3} \da_{p_4, \o_4} \ra  = - 4 i \lam_{p_1 p_2 p_3 p_4}
\sum_{i=1}^4 \hG_i \,    D_{p_1, \o_1}D_{p_2, \o_2}D_{p_3, \o_3}D_{p_4, \o_4}\, 2\pi \delta(\o_{12;34})~,
\ee
where we simply attached external propagators $D_{p_i, \o_i}$ to the vertex (\ref{quarticV2}). 
To obtain the four-point function in the time domain we can either take the Fourier transform, or compute it directly using the time domain Feynman rules. We start with a direct computation. Using the quartic interaction term (\ref{quarticInt}) and the propagator identities (\ref{411}) and (\ref{412}) in Appendix~\ref{ap:prop} gives, 
\bml
\la a_{p_1}(t_1) a_{p_2}(t_2) \da_{p_3}(t_3) \da_{p_4}(t_4) \ra \\
= -4 i\lam_{p_1 p_2 p_3 p_4} \int \!\! d t_a\, D_{p_1}(t_{a1}) D_{p_2}(t_{a2}) D_{p_3}(t_{3a}) D_{p_4}(t_{ 4a}) \(\frac{\theta(t_{1a})}{n_1} {+} \frac{\theta(t_{2a})}{n_2}  {-} \frac{\theta(t_{3a})}{n_3} {-} \frac{\theta(t_{4a})}{n_4} \)~.
\end{multline}
A special case which will be of interest later on is when all four times are equal, 
$t_1 = \ldots = t_4 = t$, 
\bml
\la a_{p_1}(t) a_{p_2}(t) \da_{p_3}(t) \da_{p_4}(t) \ra = -4 i  \lam_{p_1 p_2 p_3 p_4} \(\frac{1}{n_1} {+} \frac{1}{n_2}{-} \frac{1}{n_3}{-} \frac{1}{n_4}\) n_1 n_2 n _3 n_4 \\
\int d t_a\theta(t - t_a)  \exp\(\(- i ( \o_{p_1}{+}\o_{p_2} {-} \o_{p_3} {-}\o_{p_4}) -  \gamma_{1234}\) (t-t_a)\)~,
\end{multline}
where $\gamma_{ijkl} \equiv \gamma_{p_i} + \gamma_{p_j} + \gamma_{p_k} + \gamma_{p_l}$. The lower bound for the integration time is the initial time. As discussed previously, we are computing correlation functions at late times. Due to the dissipation, the contribution of the integral from the initial time is therefore irrelevant. Doing the integral we get, 
\be\label{treeQ}
\!\!\!\! \la a_{p_1}(t) a_{p_2}(t) \da_{p_3}(t) \da_{p_4}(t) \ra =4 \lam_{p_1 p_2 p_3 p_4}\! \(\frac{1}{n_1} {+} \frac{1}{n_2}{-} \frac{1}{n_3}{-} \frac{1}{n_4}\) n_1 n_2 n _3 n_4 \frac{1}{ \o_{p_3}{+}\o_{p_4} {-} \o_{p_1} {-}\o_{p_2} {+} i \gamma_{1234}}~.
\ee
As expected, for vanishing dissipation there is a resonance at $ \o_{p_1}{+}\o_{p_2} = \o_{p_3} {+}\o_{p_4}$.

It is instructive to recover this result from a Fourier transform of the frequency-space correlator (\ref{4ptF}), 
\be \label{4ptFT}
\la a_{p_1}(t_1) a_{p_2}(t_2) \da_{p_3}(t_3) \da_{p_4}(t_4) \ra = \int \frac{d\o_1}{2\pi} \cdots\frac{ d \o_4}{2\pi}  e^{-i (\o_1 t_1{+}\o_2 t_2{-}\o_3 t_3{-}\o_4t_4) }\la a_{p_1, \o_1} a_{p_2, \o_2} \da_{p_3, \o_3} \da_{p_4, \o_4} \ra~.
\ee
In particular, when all the times are equal, $t_1 = \ldots = t_4 = t$, we have that the equal-time four-point function is, 
\be \label{FT4}
\la a_{p_1}(t) a_{p_2}(t) \da_{p_3}(t) \da_{p_4}(t) \ra = \int \frac{d\o_1}{2\pi} \cdots\frac{ d \o_4}{2\pi}  e^{-i (\o_1{+}\o_2{-}\o_3{-}\o_4)t }\la a_{p_1, \o_1} a_{p_2, \o_2} \da_{p_3, \o_3} \da_{p_4, \o_4} \ra~.
\ee
We now evaluate the four integrals in (\ref{FT4}). First, it is clear that the answer will be time-independent, because the energy conserving delta function in (\ref{4ptF}) cancels the exponent. 
Let us look at the contribution of e.g. the $\h G_1$ term in (\ref{4ptF}). The propagator factors for $p_1$ reduce to, 
\be
\frac{(G_{p_1, \o_1}^{-1})^*}{F_{p_1}} D_{p_1, \o_1} = G_{p_1, \o_1}
\ee
where we used (\ref{DG}). We use the energy conserving delta function to perform the integral over $\o_1$. This leaves
\be
-4 i\lam_{p_1 p_2 p_3 p_4}\int \frac{d\o_2}{2\pi} \frac{d\o_3}{2\pi} \frac{ d \o_4}{2\pi}  G_{p_1, \o_3{+}\o_4{-}\o_2}D_{p_2, \o_2}D_{p_3, \o_3}D_{p_4, \o_4}~.
\ee
We perform the $\o_2$ integral, closing the contour in the lower half plane to pick up the pole at $\o_2 = \o_{p_2} - i \gamma_{p_2}$, and then we do the $\o_3$ and $\o_4$ integrals,  closing the contours in the upper half plane (of course, we could close the contours in either half plane, but the choice we gave is the simplest). We get, 
\be
4 \lam_{p_1 p_2 p_3 p_4}n_2 n_3 n_4 \frac{1}{ \o_{p_3}{+}\o_{p_4} {-} \o_{p_1} {-}\o_{p_2} {+} i \gamma_{1234}}~.
\ee
Repeating for the other three terms in parenthesis in (\ref{4ptF}), in total we end up with precisely (\ref{treeQ}). 

The computation of the tree-level six-point function is similar to that of the four-point function, and is relegated to
 Appendix~\ref{sec:43}. 

\section{One-loop diagrams} \label{sec5}
In Sec.~\ref{sec51} we compute the one-loop correction to the propagator, showing that it corresponds to a frequency shift. In Sec.~\ref{sec52} we compute the one-loop correction to the four-point function. 

\ss{Frequency renormalization} \label{sec51}

\begin{figure}
  \centering
    \includegraphics[width=0.25\textwidth]{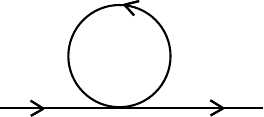}
       \caption{The one-loop correction to the propagator, (\ref{53}), which corresponds to a frequency shift. }
        \label{fig:tadpole}
\end{figure}

The one-loop correction to the propagator $D_{p_1, \o_1}$ is a simple tadpole diagram, see Fig.~\ref{fig:tadpole},
\be \label{41}
-4 i  D_{p_1, \o_1}^2 \sum_{p_2} \int \frac{d \o_2}{2\pi}  \lam_{p_1 p_2 p_2 p_1}  D_{p_2, \o_2}\(\frac{(G_{p_1, \o_1}^{-1})^* - G_{p_1, \o_1}^{-1}}{F_{p_1}} + \frac{(G_{p_2, \o_2}^{-1})^* - G_{p_2, \o_2}^{-1}}{F_{p_2}} \)~.
\ee
\!\!\!\!\!\! This is a straightforward application of the Feynman rules in Fig.~\ref{FeyRules}: there is a square of the propagator (\ref{DG})  coming from the mode $(p_1, \o_1)$ entering and leaving, and there is an interaction vertex (\ref{quarticV2}), with momentum $p_2$ running in the loop. 
We simplify the expression, noting that,
\be
\frac{(G_{p_i, \o_i}^{-1})^* - G_{p_i, \o_i}^{-1}}{F_{p_i}} = \frac{2i (\o_i - \o_{p_i})}{F_{p_i}}~,
\ee
which follows immediately from the definition of $G_{p_i, \o_i}$ in (\ref{DG}). 
Performing the integral over $\o_2$ in (\ref{41}), we find that the first term in parenthesis gives $n_2$ while the second vanishes by antisymmetry, and we get that (\ref{41}) is equal to,
\be \label{53}
2 D_{p_1, \o_1}^2  \frac{\o_1 - \omega_{p_1}}{F_{p_1}} \delta \omega_{p_1}~, \ \ \ \  \delta \omega_{p_1} = 4\sum_{p_2}   \lam_{p_1 p_2 p_2 p_1} n_2~.
\ee
We have identified this as a frequency shift, since the functional form matches what one gets from a frequency shift in the tree level propagator $D_{p_1, \o_1}$. Namely, replacing $\o_{p_1} \rightarrow \o_{p_1} + \delta \omega_{p_1}$ and expanding  $D_{p_1, \o_1}$ to leading order in $\delta \omega_{p_1}$ gives, 
\be
 D_{p_1,\o_1} \rightarrow D_{p_1,\o_1} + 2 D_{p_1,\o_1}^2\frac{\o_1 - \omega_{p_1}}{F_{p_1}}  \delta \omega_{p_1}  + \ldots~,
\ee
matching the form of (\ref{53}).

\ss{One-loop four-point function} \label{sec52}
In this section we compute the one-loop four-point function.\\

\begin{figure} \centering
\subfloat[]{\includegraphics[width=.9in]{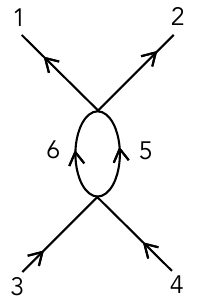}} \ \ \  \ \  \ \ 
\subfloat[]{\includegraphics[width=1.6in]{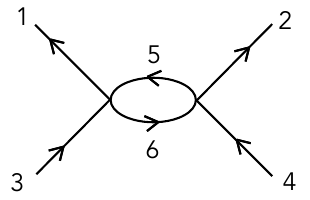}}\ \ \  \ \  \ \ 
\subfloat[]{\includegraphics[width=1.6in]{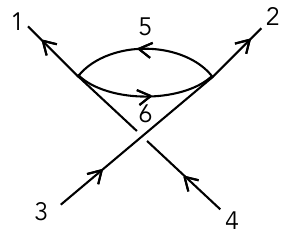}} \ \ \ \ \ \ \ \
\subfloat[]{\includegraphics[width=1.1in]{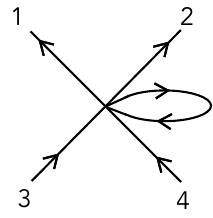}}
\caption{Feynman diagrams for the four-point function at one loop. The diagrams (a), (b), and (c) arise from the quartic interaction term. The diagram in (a)  is given by (\ref{55}). The diagram in (b) is given by (\ref{515}). The diagram in (c) is the same as diagram (b), but with $3$ and $4$ exchanged. The diagram (d) results from the sextic interaction term and is given by (\ref{511a}).} \label{Goneloop}
\end{figure}

 There are three diagrams built from the quartic vertices  and one diagram coming from the sextic vertex, see Fig.~\ref{Goneloop}. The tree-level diagram in the previous section consisted of modes $p_1$ and $p_2$ scattering directly into modes  $p_3$ and $p_4$. At one loop there are two additional modes, $p_5$ and $p_6$, and modes $p_1$ and $p_2$ either scatter into these initially (Fig.~\ref{Goneloop}(a)), or scatter off of (Fig.~\ref{Goneloop}(b) and (c)).  We now evaluate each diagram in turn. 
 
\sss*{Quartic diagrams}
We start with Fig.~\ref{Goneloop}(a). Applying the Feynman rules in Fig.~\ref{FeyRules}, it is given by, 
\be \label{55}
\!\! {-}2\pi\delta(\o_{1,2;3,4}) 8\!\!\sum_{p_5,p_6} \!\lam_{p_1 p_2 p_5 p_6}\lam_{p_5 p_6 p_3 p_4}\! \int\!\frac{d\o_5}{2\pi}(\hG_3 {+} \hG_4{+} \hG_5{+}\hG_6)(\hG_1 {+} \hG_2 {-} \hG_5^*{-}\hG_6^*)\prod_{i=1}^6 \!D_{p_i, \o_i} \Big|_{\o_6 = \o_3 {+}\o_4{-} \o_5} \ 
\ee
where  we defined $\o_{i, j; k, l} \equiv \o_i + \o_j - \o_k - \o_l$ and introduced the shorthand
\be \label{Ghat}
\hG_i = \frac{1}{G^*_{p_i, \o_i} F_{p_i}}~, \ \ \ i = 1,2,5,6~, \ \ \ \ \ \ \ \hG_i = \frac{-1}{G_{p_i, \o_i} F_{p_i}}~, \ \ \ i =3, 4~.
\ee

Upon opening the parenthesis in (\ref{55}), there are $16$ terms. We perform the $\o_5$ integral for each of the terms. Each integral can be turned into a contour integral, which is evaluated by picking up simple poles. The details are recorded in Appendix~\ref{ap:integral}. The result is, 
 \bml  \label{511v2}
{-}2\pi \delta(\o_{1,2;3,4})\, 8\sum_{p_5,p_6} \lam_{p_1 p_2 p_5 p_6}\lam_{p_5 p_6 p_3 p_4} \prod_{i=1}^4 D_{p_i, \o_i}\Big[ - \(\frac{n_6}{F_{p_5}} + \frac{n_5}{F_{p_6}} \)\\
\!\!\!\!\!+\frac{  i \o_{3,4; p_5,p_6} (n_5{+}n_6) \sum_{i=1}^4 \hG_i + \g_{56}\(  2(\hG_1{+}\hG_2)(\hG_3{+}\hG_4) n_5 n_6 +(n_5 {+} n_6)(\hG_1{ +} \hG_2 {-} \hG_3 {-} \hG_4)\) }{\o_{3,4; p_5,p_6}^2 + \g_{56}^2}\Big]
\end{multline}
where $\g_{i j} \equiv \g_{p_i} + \g_{p_j}$.

The diagram in Fig.~\ref{Goneloop}(b) is given by, 
\be  \label{515}
\!\!{-}2\pi \delta(\o_{1,2; 3,4}) 16\!\sum_{p_5,p_6}\!\! \lam_{p_1 p_6 p_3 p_5}\lam_{p_2 p_5 p_4 p_6}\! \!\int\!\frac{d\o_5}{2\pi}\!\(  \hG_1 {+}\hG_3 {-} \hG_5^*{+} \hG_6  \!\)
\!\!\( \!  \hG_2{ +}\hG_4 {+}\hG_5 {-}\hG_6^*\) \!  \prod_{i=1}^6\! D_{p_i, \o_i} \Big|_{\o_6 = \o_3 {+}\o_5{-} \o_1} \ 
\ee
Evaluating the integral over $\o_5$, using similar techniques to those we just used for Fig.~\ref{Goneloop}(a), gives, 
 \bml   \label{figb}
{-}2\pi \delta(\o_{1,2;3,4})\, 16\sum_{p_5,p_6}  \lam_{p_1 p_6 p_3 p_5}\lam_{p_2 p_5 p_4 p_6}  \prod_{i=1}^4 D_{p_i, \o_i}\Big[ - \(\frac{n_6}{F_{p_5}} + \frac{n_5}{F_{p_6}} \)\\
+\frac{  i \o_{1,p_6;3,p_5} (n_6{-}n_5) \sum_{i=1}^4\hG_i + \g_{56}\(  2(\hG_1{+}\hG_3)(\hG_2{+}\hG_4) n_5 n_6 +(n_6 {-} n_5)(\hG_{1}{-}\hG_2{+}\hG_{3}{-}\hG_4)\)}{\o_{1,p_6; 3,p_5}^2+ \g_{56}^2}\Big]
\end{multline}
The contribution of Fig.~\ref{Goneloop}(c) is the same as (\ref{figb}), but with $3 \leftrightarrow 4$. 

\sss*{Sextic diagram}
Another contribution to the four-point function is from the tadpole diagram coming from the sextic interaction, see Fig.~\ref{Goneloop}(d). We will see that  this diagram simply cancels  off the part of the contribution to the other three diagrams in Fig.~\ref{Goneloop} that is divergent in the limit that $F_{p_i}$ goes to zero; namely, the first line in (\ref{511v2}) and the first line in (\ref{figb}). 

To compute the diagram it is   convenient to rewrite the sextic interaction (\ref{sexticInt}) with $4$ and $5$ interchanged, 
\be  \label{L410}
L_{\mO(\lam^2)}  =4 \sum_{p_1, \ldots, p_7} \frac{1}{F_{p_7}} \lam_{p_1p_2p_3p_7} \lam_{p_7p_5p_4p_6} \da_{p_1}\da_{p_2} a_{p_3} a_{p_4} \da_{p_5} a_{p_6}~.
\ee
For computing the diagram in Fig.~\ref{Goneloop}(d), we Wick contract $\da_{p_5}$ with $a_{p_6}$ in $L_{\mO(\lam^2)}$ (giving $n_5$), and contract the other $a_{p_i}$ and $\da_{p_i}$ with the corresponding external legs, giving $D_{p_i, \o_i}$,~\footnote{For the rest of this subsection, for notational simplicity, $\lam_{i j k l} \equiv \lam_{p_i p_j p_k p_l}$ and $F_{i} \equiv F_{p_i}$.} 
\be \label{511a}
{-}2\pi \delta(\o_{1,2;3,4})\, 4 \prod_{i=1}^4 D_{p_i, \o_i}
\sum_{p_5,p_7} \frac{n_5}{F_7} 
\(\lam_{1237}\lam_{7545}  + \text{perm}\)~,
\ee
where we needed to also include all permutations of $(1,2,5)$ and independently of $(3,4,6)$; there are a total of $3!\times 3! = 36$ such permutations.~\footnote{Here we mean that one first includes all permutations  of $(1,2,5)$ and independently of $(3,4,6)$ for $\lam_{1237}\lam_{7546}$, and then sets $5=6$. Also, because of the symmetric footing of $1$ and $2$, and separately of $4$ and $6$, in the Lagrangian (\ref{L410}), it is most convenient to split the $36$ permutations into $9 \times 4$ permutations, where the $4$ are the identity, $(1\leftrightarrow 2)$, $(4\leftrightarrow 6)$, $(1\leftrightarrow 2, 4\leftrightarrow 6)$. }
We now use momentum conservation in the couplings so the e.g. $\lam_{2557} \lam_{7143}$ is in fact $\lam_{2525}\lam_{2143}$ because $p_7$ must be equal to $p_2$ in order for $\lam_{2557}$ to be nonzero. We also use the symmetries of the coupling $\lam_{i j k l} = \lam_{j i k l} = \lam_{i j l k}$, and change the dummy variable from $p_7$ to $p_6$.  We get
\bml
{-}2\pi \delta(\o_{1,2;3,4})\, 4 \prod_{i=1}^4 D_{p_i, \o_i}\Big[
\sum_{p_5} n_5 \lam_{1234} \sum_{i=1}^4 \frac{4 \lam_{i 55 i}}{F_i}  
\\+ \sum_{p_5, p_6} \frac{4 n_5}{F_6} \( \lam_{1256} \lam_{56 34} + \lam_{5236} \lam_{6145} + \lam_{5246} \lam_{6153} + \lam_{5136}\lam_{6245} + \lam_{5146} \lam_{6253}\)
\Big]~.
\end{multline}
We can combine the terms in the second line so as to  get,
\bml \label{523}
{-}2\pi \delta(\o_{1,2;3,4})\, 4 \prod_{i=1}^4 D_{p_i, \o_i}\Big[
\sum_{p_5} n_5 \lam_{1234} \sum_{i=1}^4 \frac{4 \lam_{i 55 i}}{F_i}  \\
+2 \sum_{p_5, p_6} \( \frac{ n_5}{F_6} + \frac{n_6}{F_5}\) \( \lam_{1256} \lam_{56 34} +2 \lam_{5236} \lam_{6145} + 2\lam_{5246} \lam_{6153}  \)
\Big]~.
\end{multline}

\sss*{Frequency space four-point function}
The four-point function is a sum of the contributions we found above  for the four diagrams in Fig.~\ref{Goneloop}.  

We start by combing the contribution of the sextic diagram (\ref{523}) with the divergent contribution (in the limit of $F_{p_i}\rightarrow 0$) coming from the quartic diagrams, Fig.~\ref{Goneloop} (a-c). For the diagram in Fig.~\ref{Goneloop}(a) this is the first line of (\ref{511v2}), for the diagram in Fig.~\ref{Goneloop}(b) this is the first line of (\ref{figb}), and for the diagram in Fig.~\ref{Goneloop}(c) this is the first line of (\ref{figb}) with $3 \leftrightarrow 4$. The sum of these three terms is,
\be \label{524}
2\pi \delta(\o_{1,2;3,4})\,  \prod_{i=1}^4 D_{p_i, \o_i}8 \sum_{p_5, p_6} \( \frac{n_5}{F_6}+ \frac{n_6}{F_5}\) \(\lam_{1256} \lam_{5634}+2\lam_{1635} \lam_{2546} +2\lam_{1645} \lam_{2536} \)~.
\ee
The sum of (\ref{523}) and (\ref{524}) is, 
\be \label{525}
\!{-}2\pi \delta(\o_{1,2;3,4})\,16\prod_{i=1}^4\! D_{p_i, \o_i}
\sum_{p_5} n_5 \lam_{p_1p_2p_3p_4} \sum_{i=1}^4 \frac{\lam_{p_i p_5p_5 p_i}}{F_{p_i}} ={-} 2\pi \delta(\o_{1,2;3,4})\,4\prod_{i=1}^4\! D_{p_i, \o_i}\lam_{p_1p_2p_3p_4}\! \sum_{i=1}^4 \frac{\delta \omega_{p_i}}{F_{p_i}}~,
\ee
where we wrote this in terms of  the frequency shift $\delta \omega_p$ in (\ref{53}). 

Eq.~\ref{525}  has a simple interpretation. Recall that in Sec.~\ref{sec51} we computed the one loop renormalization of the frequency. 
Under a renormalization of the frequency, the term in the Lagrangian transforms as $\sd_k a_k \rightarrow \sd_k a_k + i \delta\omega_k a_k$. From the cross term in (\ref{L42}) we get the term, 
\be
L \rightarrow L + \sum_{p_1, \ldots, p_4} \sum_{i=1}^4\frac{ \delta \omega_{p_i}}{F_{p_i}} \lam_{p_1 p_2 p_3 p_4} \da_{p_1}\da_{p_2} a_{p_3} a_{p_4}~.
\ee
This gives a contribution to the amplitude which precisely  matches (\ref{525}). 

Thus, the one loop four-point function is the sum of the finite (in the  limit of $F_{p_i}\rightarrow 0$) part  of Fig.\ref{Goneloop} (a-c). This consists of  the second line of (\ref{511v2}), the second line of  (\ref{figb}), and the second line of  (\ref{figb}) with $3 \leftrightarrow 4$. Combining this with the tree-level contribution found earlier in (\ref{4ptF}),  the  four-point function up to order $\lam^2$ is, 
\bml\label{4total}
\la a_{p_1, \o_1} a_{p_2, \o_2} \da_{p_3, \o_3} \da_{p_4, \o_4} \ra =  {-}2\pi \delta(\o_{1,2;3,4})\, 4i \lam_{p_1 p_2 p_3 p_4} \prod_{i=1}^4 D_{p_i, \o_i}
\sum_{i=1}^4 \hG_i \\
{-}2\pi \delta(\o_{1,2;3,4})\, \prod_{i=1}^4 D_{p_i, \o_i}
 \sum_{p_5,p_6}  \Big[ 
8 \lam_{p_1 p_2 p_5 p_6}\lam_{p_5 p_6 p_3 p_4} \\
\frac{  i \o_{3,4; p_5,p_6} (n_5{+}n_6) \sum_{i=1}^4 \hG_i + \g_{56}\(  2(\hG_1{+}\hG_2)(\hG_3{+}\hG_4) n_5 n_6 +(n_5 {+} n_6)(\hG_1{ +} \hG_2 {-} \hG_3 {-} \hG_4)\) }{\o_{3,4; p_5,p_6}^2 + \g_{56}^2}\\
\hspace{-9cm}+\Big(16 \lam_{p_1 p_6 p_3 p_5}\lam_{p_2 p_5 p_4 p_6}  \\ \frac{  i \o_{1,p_6;3,p_5} (n_6{-}n_5) \sum_{i=1}^4\!\hG_i {+} \g_{56}\(  2(\hG_1{+}\hG_3)(\hG_2{+}\hG_4) n_5 n_6 {+}(n_6 {-} n_5)(\hG_{1}{-}\hG_2{+}\hG_{3}{-}\hG_4)\)}{\o_{1,p_6; 3,p_5}^2+ \g_{56}^2}\\ + ( 3\leftrightarrow 4)\Big) \Big]~,
\end{multline}
where the first line is the tree-level contribution and the rest is the one-loop contribution.~\footnote{The frequency space four-point function is a function of the four frequencies, $\o_1, \ldots, \o_4$, whereas  the $\o_{p_i}$ (and in particular the $\o_{p_5}$ and $\o_{p_6}$ that appear here) are not   independent variables and are just  functions of $p_i$.} 
If we take the limit of $\g, F \rightarrow 0$, then we can drop terms in the numerator that are proportional to $\g$. We get the  four-point function in the $\gamma, F\rightarrow 0$ limit is, 
 \bml  
 \la a_{p_1, \o_1} a_{p_2, \o_2} \da_{p_3, \o_3} \da_{p_4, \o_4} \ra =  {-}2\pi \delta(\o_{1,2;3,4})\, 4 i \lam_{p_1 p_2 p_3 p_4} \prod_{i=1}^4 D_{p_i, \o_i}
\sum_{i=1}^4 \hG_i \\
{-}2\pi \delta(\o_{1,2;3,4})\, \prod_{i=1}^4 D_{p_i, \o_i}\Big(\sum_{i=1}^4\hG_i  \Big)
 \Big[ 
 \sum_{p_5,p_6} 8 \lam_{p_1 p_2 p_5 p_6}\lam_{p_5 p_6 p_3 p_4} 
\frac{  i \o_{3,4; p_5,p_6} (n_6{+}n_5) }{\o_{3,4; p_5,p_6}^2 + \g_{56}^2}\\
+ \Big(16 \lam_{p_1 p_6 p_3 p_5}\lam_{p_2 p_5 p_4 p_6}  \frac{  i \o_{1,p_6;3,p_5} (n_6{-}n_5)}{\o_{1,p_6; 3,p_5}^2+ \g_{56}^2} + (3\leftrightarrow 4)\Big)\ \Big]~,
\end{multline}
where the propagator $D_{p_i, \o_i}$ (\ref{DG}) in the $\g, F \rightarrow 0$ limit is $D_{p_i, \o_i}\rightarrow n_{p_i} 2\pi \delta(\o_i{-}\o_{p_i})$.

\sss*{Equal time four-point function}

The four-point function in time, $\la a_{p_1}(t_1) a_{p_2}(t_2) \da_{p_3}(t_3) \da_{p_4}(t_4) \ra $,  is simply the Fourier transform of the frequency-space four-point function found above, see (\ref{4ptFT}).  We are in particular interested in the four-point function when all the times are equal, $t_1 = \ldots = t_4 = t$. 

To compute the four-point function in time, at equal times, we take the Fourier transform. Taking the integral of (\ref{4total}) with respect to all $\o_i$, $i=1, 2, 3,4$, as prescribed by the (\ref{FT4}), we get that the equal-time four-point function  to order $\lam^2$ is, 
\bml \label{4ptloopPos}
\la a_{p_1}(t) a_{p_2}(t) \da_{p_3}(t) \da_{p_4}(t) \ra = 4\lam_{p_1 p_2 p_3 p_4}  \prod_{i=1}^4 n_i \Big(\frac{1}{n_1} {+} \frac{1}{n_2}{-} \frac{1}{n_3}{-} \frac{1}{n_4}\Big) \frac{1}{ \o_{p_3, p_4; p_1, p_2} {+} i \gamma_{1234}}\\
{-}8\sum_{p_5, p_6} \prod_{i=1}^6 n_i\, \Big\{
  \lam_{p_1 p_2 p_5 p_6}\lam_{p_5 p_6 p_3 p_4}\Big[ \Big(\frac{1}{n_1} {+} \frac{1}{n_2} \Big)\Big(\frac{1}{n_3} {+} \frac{1}{n_4} \Big) \frac{-i}{\o_{p_3, p_4; p_5, p_6}{+}i \g_{3456}}\frac{i}{\o_{p_5, p_6; p_1, p_2} {+}i \g_{1256}} \G   \Big. \\ 
\Big. +  \Big(\frac{1}{n_5} {+} \frac{1}{n_6} \Big) \( \Big(\frac{1}{n_1} {+} \frac{1}{n_2} \Big) \frac{i}{\o_{p_3, p_4; p_1, p_2}{+} i\gamma_{1234}}\frac{i}{\o_{p_3, p_4; p_5, p_6}{+} i\gamma_{3456}} + (1, 2\leftrightarrow 3,4)^* \)  \Big]\\
+\bigg(2\lam_{p_1 p_6 p_3 p_5}\lam_{p_2 p_5 p_4 p_6} 
 \Big[ \Big(\frac{1}{n_1}{-}\frac{1}{n_3}\Big)\Big(\frac{1}{n_2}{-}\frac{1}{n_4}\Big) \frac{i}{\o_{p_4,p_6; p_2, p_5} {+}i \g_{2456}}
\frac{i}{\o_{p_3,p_5; p_1, p_6} {+}i\g_{1356}}\G
\\
 +\Big(\frac{1}{n_5} {-} \frac{1}{n_6} \Big) \(\Big(\frac{1}{n_1}{ -}\frac{1}{n_3}\Big)\frac{i}{\o_{p_3, p_4; p_1, p_2}{+} i\gamma_{1234}}\frac{i}{\o_{p_4,p_6; p_2, p_5}{+} i\gamma_{2456}} - (1 \leftrightarrow 2, 3\leftrightarrow 4, 5\leftrightarrow 6)\) \Big]  \\ + \( 3 \leftrightarrow 4\) \bigg) \Big\}~,
\end{multline}
where 
\be
\G = 1+\frac{2 \g_{56} i}{\o_{p_3, p_4; p_1, p_2}{+}i \g_{1234}} ~.
\ee
where the first line is the tree-level contribution found earlier in  (\ref{treeQ}). \\

\noindent \textit{Vanishing forcing and dissipation}
\\[-5pt]

Our result (\ref{4ptloopPos}) is valid with finite forcing and finite dissipation. Let us look at it 
in the limit in which both forcing and dissipation vanish (while maintaining a constant ratio; $\g_k, F_k\rightarrow 0$, $n_k \equiv F_k/(2\g_k) =\text{const.}$). 

We may drop factors of $\g$ in the numerator in (\ref{4ptloopPos}), which  simply sets $\G = 1$.~\footnote{This is not correct for $\o_{p_3, p_4; p_1, p_2}=0$, however it is correct when viewed as a distribution, which is what is relevant. We thank D.~Schubring for discussions on this point. } We write the four-point function   in a compact form which makes its symmetries manifest,
\be \label{4ptss}
\la a_{p_1}(t) a_{p_2}(t) \da_{p_3}(t) \da_{p_4}(t) \ra =\la a_{p_1}(t) a_{p_2}(t) \da_{p_3}(t) \da_{p_4}(t) \ra_{\text{tree}} + \la a_{p_1}(t) a_{p_2}(t) \da_{p_3}(t) \da_{p_4}(t) \ra_{\text{one-loop}}~,
\ee
where the tree-level contribution is,
\be
\la a_{p_1}(t) a_{p_2}(t) \da_{p_3}(t) \da_{p_4}(t) \ra_{\text{tree}}  =  4\lam_{p_1 p_2 p_3 p_4}  \prod_{i=1}^4 n_i \Big(\frac{1}{n_1} {+} \frac{1}{n_2}{-} \frac{1}{n_3}{-} \frac{1}{n_4}\Big) \frac{1}{ \o_{p_3, p_4; p_1, p_2} + i\eps}
\ee
and the one-loop contribution is, 
\bml \label{424}
 \la a_{p_1}(t) a_{p_2}(t) \da_{p_3}(t) \da_{p_4}(t) \ra_{\text{one-loop}} 
 =8 \sum_{p_5, p_6}\Big( \lam_{p_1 p_2 p_5 p_6}\lam_{p_5 p_6 p_3 p_4} V(1,2,3,4,5, 6) \\- 2\lam_{p_1 p_6 p_3 p_5}\lam_{p_2 p_5 p_4 p_6} V(1,-3,-2,4,5,-6)
-2 \lam_{p_1 p_6 p_4 p_5}\lam_{p_2 p_5 p_3 p_6} V(1,{-}4,{-}2,3,5,{-}6)\Big)
 \end{multline}
 where we defined, 
 \bml
 V(1,2,3,4,5,6) =
 -\prod_{i=1}^6 n_i \Big\{\Big(\frac{1}{n_1} {+} \frac{1}{n_2} \Big)\Big(\frac{1}{n_3} {+} \frac{1}{n_4} \Big) \frac{1}{\o_{p_3, p_4; p_5, p_6}{+}i\eps}\frac{1}{\o_{p_5, p_6; p_1, p_2} {+}i \eps} \\- \Big(\frac{1}{n_5} {+} \frac{1}{n_6} \Big)\Big(\frac{1}{n_1} {+} \frac{1}{n_2} \Big) \frac{1}{\o_{p_3, p_4; p_1, p_2}{+}i\eps}\frac{1}{\o_{p_3, p_4; p_5, p_6} {+}i \eps} - \Big(\frac{1}{n_5} {+} \frac{1}{n_6} \Big)\Big(\frac{1}{n_3} {+} \frac{1}{n_4} \Big) \frac{1}{\o_{p_3, p_4; p_1, p_2}{+}i\eps}\frac{1}{\o_{p_5, p_6; p_1, p_2} {+}i \eps}  \Big\}
 \end{multline}
Note that in (\ref{424}) when we write one of the numbers in $V(...)$ to have a minus sign we mean that both the corresponding $n_i$ and $\o_{p_i}$ come with a minus sign. 
If we wish, we may write (\ref{424}) in a more compact form, 
\bml
\la a_{p_1}(t) a_{p_2}(t) \da_{p_3}(t) \da_{p_4}(t) \ra_{\text{one-loop}} \\
 =8 \sum_{p_5, p_6} \[\lam_{p_1 p_2 p_5 p_6}\lam_{p_5 p_6 p_3 p_4} V(1,\ldots, 6) - 2(2\leftrightarrow {-}3, 6\rightarrow -6) -  2 (2\leftrightarrow -4, 6\rightarrow -6)\]~,
  \end{multline}
  where when applying the permutation on the couplings, such as $ (2\leftrightarrow -3, 6\rightarrow -6)$, we have that e.g., 
  \be
  \lam_{p_1 p_2 p_5 p_6} \rightarrow \lam_{p_1 -p_3 p_5 - p_6} \equiv \lam_{p_1 p_6 p_5 p_3}~,
  \ee
  and recall that $\lam_{p_1 p_6 p_5 p_3} = \lam_{p_1 p_6 p_3 p_5}$. Now, we will use that 
\be
\frac{1}{x+i\eps}  = \frac{1}{x} - i \pi \delta(x)~,
\ee
where $1/x$ means the principal value of $1/x$. This kind of splitting is useful, because $1/x$ is odd under $x\rightarrow - x$ while $\delta(x)$ is even. We get that, 
\be
V(1, 2,3,4,5,6)  = V_{pp}(1, 2,3,4,5,6) + i V_{p\delta}(1, 2,3,4,5,6) + V_{\delta \delta}(1, 2,3,4,5,6)~.
\ee
where
\bea  
V_{pp}(1, 2,3,4,5,6) &=& A_{pp}(1, 2,3,4,5,6) +  A_{pp}(3,4,5,6,1,2) +  A_{pp}(5,6,1,2,3,4)   \nn
\\ V_{p\delta}(1, 2,3,4,5,6) &=& A_{p\delta}(1, 2,3,4,5,6) -  A_{p\delta}(3,4,5,6,1,2) - A_{p\delta}(5,6,1,2,3,4)\nn
\\V_{\delta\delta}(1, 2,3,4,5,6) &=&  A_{\delta\delta}(1, 2,3,4,5,6) -  A_{\delta\delta}(3,4,5,6,1,2) -  A_{\delta\delta}(5,6,1,2,3,4)~, \label{429}
\eea
where
\bea \nn
A_{pp}(1, 2,3,4,5,6) &=&-\prod_{i=1}^6 n_i\Big(\frac{1}{n_1} {+} \frac{1}{n_2} \Big)\Big(\frac{1}{n_3} {+} \frac{1}{n_4} \Big) \frac{1}{\o_{p_3, p_4; p_5, p_6}}\frac{1}{\o_{p_5, p_6; p_1, p_2} } \\
A_{\delta\delta}(1, 2,3,4,5,6) &=&\pi^2\prod_{i=1}^6 n_i \Big(\frac{1}{n_1} {+} \frac{1}{n_2} \Big)\Big(\frac{1}{n_3} {+} \frac{1}{n_4} \Big) \delta(\o_{p_3, p_4; p_5, p_6})\delta(\o_{p_5, p_6; p_1, p_2} ) \\ \nn
A_{p\delta}(1, 2,3,4,5,6) &=&- \pi\prod_{i=1}^6 n_i \Big(\frac{1}{n_1} {+} \frac{1}{n_2}{-}\frac{1}{n_3} {-} \frac{1}{n_4} \Big)\Big(\frac{1}{n_5} {+} \frac{1}{n_6} \Big) \frac{1}{\o_{p_3, p_4; p_5, p_6}}\delta(\o_{p_1, p_2; p_3, p_4} )~.\label{As}
\eea
Notice that $A_{pp}$ and $A_{\delta \delta}$ are even under interchange $1,2 \leftrightarrow 3,4$, whereas $A_{p \delta}$ is odd. Explicitly, $A_{pp}(1, 2,3,4,5,6) =  A_{pp}(3,4,1,2,5,6) $, while $A_{p\delta}(1, 2,3,4,5,6) = - A_{p\delta}(3,4,1,2,5,6) $.

\ss{Kinetic Equation} \label{sec53}

The kinetic equation governs the behavior of the expectation value of the mode number operator at time $t$, $n_k(t) = \la \da_k(t) a_k(t)\ra$. This is simply a two-point function, so we can compute it using the same Feynman  diagram methods as we used for the four-point function. To not do any additional work, we note that, in the limit of vanishing viscosity, one can use the equations of motion to express the expectation value of $n_k(t)$ in terms of the imaginary part of the four-point function, see (\ref{Qkinetic_eq})  in Appendix~\ref{appx:quartic}, 
\be \label{424v2}
 {\partial n_k(t)\over \partial t}     = -4 \, \text{Im} \Big(\sum_{p_i} \delta_{k  p_1} \lam_{p_3 p_4 p_1 p_2} \la a_{p_1}(t) a_{p_2}(t) \da_{p_3}(t) \da_{p_4} (t)\ra  \Big)~,
\ee
where we took the complex conjugate of (\ref{Qkinetic_eq}), and recall that $ \lam^*_{p_1 p_2 p_3 p_4} = \lam_{p_3 p_4 p_1 p_2}$. 
This equation is useful because it gives us the two-loop occupation number from the one-loop four-point function, and one-loop amplitudes are easier to compute. An alternative and equivalent form of (\ref{424v2}) is, 
\be
 {\partial n_k(t)\over \partial t}     =i \sum_{p_i} (\delta_{k  p_1} {+} \delta_{k  p_2} {-} \delta_{k  p_3} {-}\delta_{k  p_4} ) \lam_{p_3 p_4 p_1 p_2} \la a_{k}(t) a_{p_2}(t) \da_{p_3}(t) \da_{p_4} (t)\ra~.
\ee
We will stick with (\ref{424v2}). All we need to do is insert the four-point function into this expression. We will use the four-point function we computed in the previous section,  (\ref{4ptss}) which is valid up to order $\lam^2$. We get, 
\be
 {\partial n_k(t)\over \partial t}  =  \({\partial n_k(t)\over \partial t}\)_{\text{tree}} +  \({\partial n_k(t)\over \partial t} \)_{\text{one-loop}} + \ldots~.
 \ee
 where
 \be \label{434t}
  \({\partial n_k(t)\over \partial t}\)_{\text{tree}} = 16 \sum_{p_1, \ldots, p_4}\delta_{k, p_1}| \lam_{p_1 p_2 p_3 p_4}|^2  \prod_{i=1}^4 n_i\, \Big( \frac{1}{n_1} {+} \frac{1}{n_2}{-}\frac{1}{n_3} {-} \frac{1}{n_4} \Big) \, \pi \delta(\o_{p_1,p_2; p_3, p_4})
\ee
is the standard and well-known result, and the first correction is, 
  \be
  \({\partial n_k(t)\over \partial t}\)_{\text{one-loop}}  \equiv   \({\partial n_k(t)\over \partial t}\)_{\text{one-loop}}^a +   \({\partial n_k(t)\over \partial t}\)_{\text{one-loop}}^b~, 
    \ee
  where
\bml \label{kins}
  \({\partial n_k(t)\over \partial t}\)_{\text{one-loop}}^a  = -32\sum_{p_1, \ldots, p_6}\delta_{k, p_1}
   \Big[\text{Re}( \lam_{p_3 p_4 p_1 p_2} \lam_{p_1 p_2 p_5 p_6}\lam_{p_5 p_6 p_3 p_4} ) V_{p \delta}(1,2,3,4,5,6) \\ + \text{Im}( \lam_{p_3 p_4 p_1 p_2} \lam_{p_1 p_2 p_5 p_6}\lam_{p_5 p_6 p_3 p_4} ) ( V_{\delta \delta}(1,2,3,4,5,6)+V_{pp}(1,2,3,4,5,6)) \Big]~, 
\end{multline}
where $V_{pp}$, $V_{p \delta}$, and $V_{\delta \delta}$ where given earlier in (\ref{429}), and 
\bml \label{kint}
  \({\partial n_k(t)\over \partial t}\)_{\text{one-loop}}^ b= 128\sum_{p_1, \ldots, p_6}\delta_{k, p_1} \Big[\text{Re}( \lam_{p_3 p_4 p_1 p_2}\lam_{p_1 p_6 p_3 p_5}\lam_{p_2 p_5 p_4 p_6}) V_{p \delta}(1,-3,-2,4,5,-6) \\ + \text{Im}( \lam_{p_3 p_4 p_1 p_2}\lam_{p_1 p_6 p_3 p_5}\lam_{p_2 p_5 p_4 p_6}) ( V_{\delta \delta}(1,-3,-2,4,5,-6) +V_{pp}(1,-3,-2,4,5,-6) ) \Big]~,
\end{multline}
which is (\ref{kins}) with $ (2\leftrightarrow -3, 6\rightarrow -6)$ times a combinatorial factor of $4$. Note that the last two of the three contributions in (\ref{424}) give the same contribution to  (\ref{kint}), because they are related by an exchange $3\leftrightarrow 4$ of the dummy variables being summed over. 

This form of the answer is satisfactory, but we can actually simplify further. This is done in Appendix~\ref{apE} and we find, 
\bml \label{s439o}
  \({\partial n_k(t)\over \partial t}\)_{\text{one-loop}}^a  = -4^3\sum_{p_1, \ldots, p_6}\delta_{k, p_1} 
   \Big[\text{Re}( \lam_{p_3 p_4 p_1 p_2} \lam_{p_1 p_2 p_5 p_6}\lam_{p_5 p_6 p_3 p_4} ) A_{p \delta}(1,2,3,4,5,6) \\ + \text{Im}( \lam_{p_3 p_4 p_1 p_2} \lam_{p_1 p_2 p_5 p_6}\lam_{p_5 p_6 p_3 p_4} )A_{\delta \delta}(1,2,3,4,5,6) \Big]~,
\end{multline}
as well as, 
\bml
  \({\partial n_k(t)\over \partial t}\)_{\text{one-loop}}^b  =4^4
\sum_{p_1, \ldots, p_6}\delta_{k, p_1} 
   \Big[\text{Re}( \lam_{p_3 p_4 p_1 p_2} \lam_{p_1 p_6 p_3 p_5}\lam_{p_2 p_5 p_4 p_6} )A_{p \delta}(1,-3,-2,4,5,-6) \\ + \text{Im}( \lam_{p_3 p_4 p_1 p_2} \lam_{p_1 p_6 p_3 p_5}\lam_{p_2 p_5 p_4 p_6} )A_{\delta \delta}(1,-3,-2,4,5,-6) \Big]~,
\end{multline}
where $A_{p\delta}$ and $A_{\delta \delta}$ where given in (\ref{As}). 

We note that the thermal solution, $n_k \sim 1/\o_k$ is manifestly a stationary solution of the kinetic equation. For the tree-level term (\ref{434t}), one gets $(\o_1 {+}\o_2 {-} \o_3 {-} \o_4) \delta(\o_1 {+}\o_2 {-} \o_3 {-} \o_4)$, which vanishes. For the same reason, the $A_{p\delta}$ term in (\ref{s439o}) also vanishes. The vanishing of the $A_{\delta \delta}$ term is because $A_{\delta \delta}(1,2,3,4,5,6)$ goes like $(\o_{p_1}{+}\o_{p_2})^2\delta(\o_{p_3, p_4; p_5, p_6})\delta(\o_{p_5, p_6; p_1, p_2} )$, which is symmetric under interchange of $(3,4)\leftrightarrow (5,6)$, whereas the imaginary part of the coupling is antisymmetric. 

Writing everything together on one line we finally have, 
\bml \label{441}
{\partial n_k(t)\over \partial t} =16 \sum_{p_1, \ldots, p_4}\delta_{k, p_1} \pi \delta(\o_{p_1,p_2; p_3, p_4})| \lam_{p_1 p_2 p_3 p_4}|^2  \prod_{i=1}^4 n_i\, \Big( \frac{1}{n_1} {+} \frac{1}{n_2}{-}\frac{1}{n_3} {-} \frac{1}{n_4} \Big) \,\\
+ 64\sum_{p_1, \ldots, p_6} \delta_{k, p_1}  \pi \delta(\o_{p_1, p_2; p_3, p_4} )\prod_{i=1}^6  n_i \,
   \Big\{ \frac{\text{Re}( \lam_{p_3 p_4 p_1 p_2} \lam_{p_1 p_2 p_5 p_6}\lam_{p_5 p_6 p_3 p_4} ) }{\o_{p_3, p_4; p_5, p_6}} \Big(\frac{1}{n_1} {+} \frac{1}{n_2}{-}\frac{1}{n_3} {-} \frac{1}{n_4} \Big)\Big(\frac{1}{n_5} {+} \frac{1}{n_6} \Big)\\ + \text{Im}( \lam_{p_3 p_4 p_1 p_2} \lam_{p_1 p_2 p_5 p_6}\lam_{p_5 p_6 p_3 p_4} )\pi \delta(\o_{p_3, p_4; p_5, p_6} )\Big(\frac{1}{n_1} {+} \frac{1}{n_2} \Big)\Big(\frac{1}{n_3} {+} \frac{1}{n_4} \Big)\\
+4  \Big[    \frac{\text{Re}( \lam_{p_3 p_4 p_1 p_2} \lam_{p_1 p_6 p_3 p_5}\lam_{p_2 p_5 p_4 p_6} ) }{\o_{p_4, p_6; p_2, p_5}} \Big(\frac{1}{n_1} {+} \frac{1}{n_2}{-}\frac{1}{n_3} {-} \frac{1}{n_4} \Big)\Big(\frac{1}{n_5} {-} \frac{1}{n_6} \Big)
    \\ + \text{Im}( \lam_{p_3 p_4 p_1 p_2} \lam_{p_1 p_6 p_3 p_5}\lam_{p_2 p_5 p_4 p_6} ) \pi \delta(\o_{p_4, p_6; p_2, p_5}) \Big(\frac{1}{n_1} {-} \frac{1}{n_3} \Big)\Big(\frac{1}{n_2} {-} \frac{1}{n_4} \Big)\ \Big] \Big\}~.
\end{multline}
The coupling is in many cases real. If this is the case we can write the kinetic equation, up to order $\lam^3$, as
\begin{multline}   \label{11}
{\partial n_k(t)\over \partial t}  = 16\pi \sum_{p_2,\ldots, p_4} \delta_{k, p_1}  \delta(\o_{p_1,p_2; p_3 p_4}) \lam_{p_1p_2p_3p_4}^2  \prod_{i=1}^4 n_i\, \Big( \frac{1}{n_1} {+} \frac{1}{n_2}{-}\frac{1}{n_3} {-} \frac{1}{n_4} \Big)\\
\[ 1 + 4 \sum_{p_5,p_6} \frac{\lam_{p_1p_2p_5p_6}\lam_{p_5p_6p_3p_4}}{\lam_{p_1p_2p_3p_4}} \frac{n_5 {+} n_6}{\o_{p_1 p_2;p_5 p_6}}+ 16\sum_{p_5,p_6} \frac{\lam_{p_1p_6p_3p_5}\lam_{p_2p_5p_4p_6}}{\lam_{p_1p_2p_3p_4}} \frac{n_6 {-} n_5}{\o_{p_4 p_6;p_2 p_5}}\]~.
\end{multline}

This is the kinetic equation governing the mode occupation number $n_k(t)$ for mode $k$, in the limit of vanishing forcing and vanishing dissipation. 
The order $\lam^2$ term (the first line, coming from the tree-level four-point function) is the standard kinetic equation in wave turbulence, see (\ref{Q315}) in Appendix~\ref{appx:quartic}. Note that, the $n_i \equiv n_{k_i}$ on the right-hand side was defined to be $F_{k_i}/2\g_{k_i}$ (it results from its appearance in the propagator (\ref{prop})). At leading order in the coupling (i.e. for the free theory), $n_i$ corresponds to the mode occupation number, because the propagator $D_{k_i}(0) \equiv \la \da_{k_i}(t) a_{k_i}(t)\ra = n_{k_i}$, which explains our use of this notation. Also, recall that with our notation we defined $\o_{p_i, p_j; p_k,p_l} \equiv\o_{p_i} {+} \o_{p_j}{-}\o_{p_k}{-}\o_{p_l}$.

The order $\lam^2$ term in the kinetic equation has the usual interpretation: there is some loss of occupation number of  mode $k$  when mode $k$ scatters off of mode $p_2$ producing modes $p_3$ and $p_4$. Conversely, the reverse process can happen, leading to production of mode $k$. These processes are captured by the tree-level Feynman diagram Fig.~\ref{FeyRules} (b). 

The other terms in the kinetic equation, the order $\lam^3$ contribution (coming from the one-loop four-point function) is the first correction, found earlier in \cite{Polyakov, Gurarie, Gurarie95} for the special case of real couplings. The description of these processes are captured by the Feynman diagrams Fig.~\ref{Goneloop} (a-c), and involve two additional intermediate modes $k_5$ and $k_6$ in the process.

\section{Discussion} \label{sec7}

Let us summarize, first at a technical level: we have taken a classical field theory with an arbitrary, but small, quartic interaction (a collection of coupled harmonic oscillators) with dissipation and Gaussian-random forcing (\ref{eom1}, \ref{Pf2}) and have given a prescription for computing correlation functions, perturbatively in the coupling. We applied this to compute the two-point and four-point correlation functions, to next-to-leading order in the coupling.  In particular, the two-point function at leading order in the coupling was given in (\ref{prop38}) and in (\ref{DG}), and at  next-to-leading order  in (\ref{53}). Taking the two times in the two-point function to be equal gives the expectation value of the occupation number of mode $k$, which is the quantity most commonly studied (at leading order in the coupling).
The frequency-space four-point function up to next-to-leading order in the coupling was given in (\ref{4total}). The Fourier transform of the frequency space correlation function gives the correlation function with fields at different times. In the special case that all four times are equal, the equal-time four-point function to next-to-leading order was given in (\ref{4ptloopPos}). 

In the limit of vanishing forcing and vanishing dissipation, one might have thought that the properties of the system should be the same as those of a closed system, whose late time state is the thermal state. This is not so. As reviewed in Appendix~\ref{appx:quartic}, if the interactions and  dispersion relation take the form of homogeneous functions (in the mathematical sense), there is a choice of $n_k\equiv F_k/2\g_k$ (kept finite in the limit of  vanishing $F_k,\g_k$) which, at leading order, gives an alternate stationary state -- the Kolmogorov-Zakharov state (the turbulent state), which has a flux of energy passing through. With this choice of $n_k$, our equations for the correlation functions at next-to-leading order characterize fluctuations about the Kolmogorov-Zakharov  state. This includes the next-to-leading order correction to the Kolmogorov-Zakharov  state itself, as characterized by the next-to-leading order kinetic equation (\ref{441}).

 There are two broad goals which motivate  study of correlation functions in the turbulent state. On the one hand, this is a stationary state which is not the thermal state (and is relevant even in contexts of closed systems with far-from-equilibrium initial conditions) and one would like to develop linear response theory for it. For instance, one would like to find the turbulent-state analog of transport coefficients and the fluctuation-dissipation theorem, concepts familiar for the thermal state. Independently, there has been enormous interest in recent years in the study of many-body chaos. Viewing wave turbulence in light of these new developments may be productive.

 It is straightforward to apply the same methods developed here to the computation of correlation functions to higher order in the coupling, or to the computation of higher-point correlation functions, or to the case of cubic interactions instead of quartic interactions. Likewise, one may compute information-theoretic measures, such as e.g.~entanglement entropy in mode space  \cite{FalkovichShavit}. The statistics of the field $a_k$ is governed by some probability distribution. From the correlation functions one can determine this distribution, perturbatively in the coupling. In fact, since we have correlation functions with fields sitting at different times, we can say more than this, and discuss the dynamical properties of the turbulent state. 
  
 The next-to-leading order corrections to the correlation functions that we computed should, in principle, be measurable quantities. This may be a fruitful avenue to pursue, given the extensive recent experimental work on wave turbulence  \cite{FalconMordant, Cortet, Falcon, Falcon2} and prethermalization \cite{ Erne:2018gmz, Prufer, Glidden:2020qmu}. 
   
  It will  be useful to generalize the methods in this paper to wave turbulence in quantum field theory. Wave turbulence in the quark-gluon plasma produced in heavy ion collisions \cite{Schlicting,Berges:2013eia, Schlichting:2019abc} and wave turbulence in reheating in the early universe \cite{Micha:2004bv,Amin:2014eta, Chatrchyan:2020cxs } are two cases of clear experimental relevance. More generally, while thermal field theory is by now a well-established part of quantum field theory, far-from-equilibrium quantum field theory is one of the new frontiers. There has been no clear unifying theoretical framework, and progress has been a patchwork of directions and results (e.g quenches \cite{Calabrese:2007rg, Calabrese:2016xau}). We believe wave turbulence in quantum field theory will be one such fruitful direction, and of relevance to practitioners of both high energy  and condensed matter physics.

\sss*{Acknowledgments} 
We thank G.~Falkovich
and V.~Oganesyan
for helpful discussions. The work of VR is supported in part by NSF grant 2209116. 
The work of MS is supported by the  Israeli Science Foundation Center of Excellence (grant No.~2289/18) and the Quantum Universe I-CORE program of the Israel Planning and Budgeting Committee (grant No.~1937/12)

\appendix

\sec{Wave turbulence}
\label{appx:quartic}
In this Appendix we briefly review some aspects of wave turbulence \cite{Falkovichsummary, Falkovich, Nazarenko}.\\

Turbulence is a state of a nonlinear system in which energy is distributed over many degrees of freedom, in a fashion which strongly deviates from equilibrium, exhibiting chaos in both space and time \cite{Falkovichsummary}. The simplest and most tractable context in which to study turbulence is for weakly interacting waves. This is referred to as wave turbulence (or weak turbulence).~\footnote{Turbulence was originally, and is more commonly, discussed in the context of hydrodynamics -- Navier-Stokes equation.  This is not wave turbulence -- there is no weak coupling expansion, and the problem of understanding hydrodynamic turbulence, and the associated range of nonperturbative phenomena, is  difficult.  The distribution of energy per mode in the inertial regime was postulated by Kolmogorov assuming universality and scale invariance. Deriving the Kolmogorov spectrum, as well as other correlation functions which are known to differ from Kolmogorov scaling, is a long-standing  problem. In contrast, in wave turbulence, weak coupling allowed Zakharov to easily find the distribution of energy per mode (the Kolmogorov-Zakharov) spectrum. And, as this paper emphasizes, finding correlation functions more generally is straightforward. } 

In fact, wave turbulence often refers to something more specific: since the interactions are weak, one can explicitly find the kinetic equation describing the occupation numbers of each wave mode. It was demonstrated by Zakharov \cite{Zakharov} that for certain forms of the dispersion relation and interaction, there is a stationary, far-from-equilibrium, solution of the kinetic equations -- the Kolmogorov-Zakharov (KZ) solution. Wave turbulence occurs in an incredible range of physical contexts, such as: gravity and capillary surface waves, sound waves, Alfven waves, plasma waves, internal waves, nonlinear optics, Bose-Einstein condensates, and gravitational waves. 
\\

\textit{Forced and freely decaying turbulence:} There are two kinds of turbulent cascades which are usually discussed: forced turbulence and freely decaying turbulence.~\footnote{In the context of surface gravity waves, see \cite{Zakharov2} for a discussion of forced turbulence and \cite{Zakharov1} for a discussion of freely decaying turbulence.} In forced turbulence the system is driven at long wavelengths and, due to interactions, the energy cascades to short wavelength modes. At sufficiently short wavelengths the energy is dissipated. The setup is such that the system is stationary, while still being far from equilibrium. In the context of freely decaying turbulence, one has a closed Hamiltonian system and starts with far-from-equilibrium initial conditions in which, for instance, the energy is concentrated in long wavelength modes. As time evolves, energy cascades to the shorter wavelength modes, with equilibrium reached at very late times. If there is a separation of scales, then at intermediate times there will be a turbulent cascade, which is approximately steady state and the same as the one found in forced turbulence. \\

\textit{Averaging:}
Turbulent cascades are present in  interacting, chaotic, many-body systems;  to perform calculations, or measurements, some kind of averaging is necessary. Assuming statistical spatial homogeneity, one can average over initial conditions, or alternatively one can perform a time average. Alternatively,  in the context of forced turbulence, one can average over the forcing function. The expectation value $\la \cdots \ra$ will denote averaging over one of these quantities. For deriving the leading order kinetic equation for weak turbulence, which kind of averaging is used is largely irrelevant. At leading order one can simply assume higher-point correlation functions factorize into two-point functions; this is referred to as the random phase approximation, and is what we use in this appendix. The approach of introducing forcing is what is done in the main body of the text; its advantage is that it gives a systematic way of going to higher order in the coupling.

\ss{The kinetic equation}

\sss*{The leading order kinetic equation}
We give the standard derivation of the kinetic equation describing the occupation number of mode $k$ for a weakly nonlinear system with quartic interaction. For certain forms of the dispersion relation and certain interactions, the kinetic equation has a stationary solution, which is known as  the Kolmogorov-Zakharov (KZ) solution. 
The kinetic equation governs   the occupation number $n_k(t)  = \la \da_k(t) a_k(t)\ra$ at late times.~\footnote{In this appendix we define $n_k$ as the occupation number, $n_k(t)  = \la \da_k(t) a_k(t)\ra$, whereas previously in (\ref{nk}) we defined it as $n_k = F_k/2\g_k$; at leading order, they are the same.}

The Hamiltonian in terms of complex amplitudes $a_k$ was given in \reef{H21} and the equations of motion were given in (\ref{eom}), 
\be \label{Q33}
i \frac{\d a_k}{\d t} = \omega_k a_k + 2 \sum_{p_2,p_3, p_4} \lam_{k p_2 p_3 p_4} \da_{p_2} a_{p_3} a_{p_4} ~.
\ee
Multiplying \reef{Q33} by $a_k^\dagger$, taking the expectation value over the ensemble of initial conditions and subtracting the complex conjugate of the same expression, yields the kinetic equation,
\be
\label{Qkinetic_eq}
 {\partial n_k\over \partial t}     = 4\, \text{Im} \Big( \sum_{p_2, p_3, p_4} \lam_{k p_2 p_3 p_4} \la \da_k \da_{p_2} a_{p_3} a_{p_4} \ra  \Big)  ~,
\ee
where $ \la \da_k \da_{p_2} a_{p_3} a_{p_4} \ra $ is really $ \la \da_k(t) \da_{p_2}(t) a_{p_3} (t)a_{p_4} (t)\ra$. 
To find the right-hand side we first compute $ \frac{\d}{\d t} \la \da_k \da_{p_2} a_{p_3} a_{p_4} \ra$ by using the equations of motion (\ref{Q33}). We get, 
\bml
\!\!\!\!\!\!\!\( i \frac{\d}{\d t} +(\omega_k{+} \omega_{p_2}{ -} \omega_{p_3}{-} \omega_{p_4})\)   \la \da_k \da_{p_2} a_{p_3} a_{p_4} \ra =   - 2  \sum_{q_2,q_3, q_4} \lam^*_{k q_2 q_3 q_4} \la a_{q_2} \da_{q_3} \da_{q_4} \da_{p_2} a_{p_3} a_{p_4} \ra +(k\leftrightarrow p_2) \\
+ 2  \sum_{q_2,q_3, q_4} \lam_{p_3 q_2 q_3 q_4} \la \da_{q_2} a_{q_3} a_{q_4} \da_k \da_{p_2} a_{p_4} \ra
+ (p_3\leftrightarrow p_4) ~.
\label{eom4p}
\end{multline}
To leading order, the state of the wave system  is Gaussian and couples $a_k$ to $\da_k$ only. Hence, the standard Wick contraction can be used to calculate the six point functions in the above expression, {\it e.g.,} all possible contractions of $a_{q_2}$ in the first correlator on the right hand side take the form,
\bea
\la a_{q_2} \da_{q_3} \da_{q_4} \da_{p_2} a_{p_3} a_{p_4} \ra &=&  
 \la a_{q_2} \da_{q_3}\ra \la \da_{q_4} \da_{p_2}  a_{p_3} a_{p_4} \ra 
 + \la a_{q_2} \da_{q_4}\ra \la \da_{q_3} \da_{p_2}  a_{p_3} a_{p_4} \ra  
 \nonumber \\
&+& \la a_{q_2} \da_{p_2}\ra \la \da_{q_3}  \da_{q_4} a_{p_3}   a_{p_4} \ra  + \mathcal{O}(\lambda)~.
\eea
Substituting into \reef{eom4p} and using the symmetries of $\lam_{k q_2 q_3 q_4}$ yields,
\bea
&&\( i \frac{\d}{\d t} +(\omega_k{+} \omega_{p_2}{ -} \omega_{p_3}{-} \omega_{p_4})\)   
 \la \da_k \da_{p_2} a_{p_3} a_{p_4} \ra 
= (-\delta\omega_k - \delta\omega_{p_2} + \delta\omega_{p_3} + \delta\omega_{p_4}) \la \da_k \da_{p_2} a_{p_3} a_{p_4} \ra 
\nonumber \\
&-& 2\sum_{q_2,q_3, q_4} \lam^*_{k q_2 q_3 q_4} \la a_{q_2} \da_{p_2}\ra \la \da_{q_3}  \da_{q_4} a_{p_3}   a_{p_4} \ra +(k\leftrightarrow p_2) 
\nonumber \\
&+& 2  \sum_{q_2,q_3, q_4} \lam_{p_3 q_2 q_3 q_4} \la \da_{q_2} a_{p_4} \ra \la a_{q_3} a_{q_4} \da_k \da_{p_2}  \ra
+ (p_3\leftrightarrow p_4) + \mathcal{O}(\lambda^2)~,
\eea
where the frequency shifts arise from self-contractions (we computed them in Sec.~\ref{sec52}, see Eq.~\ref{53}). Redefining the frequencies to absorb the shift,  $\o_k \rightarrow \o_k + \delta \omega_k$, and carrying out the Wick contractions to calculate the four-point functions and using the definition of occupation numbers, we obtain at leading order in $\lam$, 
\bea \nn
&&\( i \frac{\d}{\d t} +(\omega_k{+} \omega_{p_2}{ -} \omega_{p_3}{-} \omega_{p_4})\)   
 \la \da_k \da_{p_2} a_{p_3} a_{p_4} \ra \\
&=& \nn
\(- 4 \lam^*_{k p_2 p_3 p_4} n_{p_2} n_{p_3}n_{p_4} +(k\leftrightarrow p_2)  \)
+\( 4  \lam^*_{k p_2 p_3 p_4} n_k n_{p_2} n_{p_4} 
+ (p_3\leftrightarrow p_4) \)~ \\
&=&
 - 4 \lam^{*}_{k p_2 p_3 p_4} \({1\over n_k} {+} {1\over n_{p_2}}{ -} {1\over n_{p_3}} {-} {1\over n_{p_4}}  \)  
n_k n_{p_2} n_{p_3}n_{p_4} ~.
\eea

Solving the differential equation, using that the right-hand side is constant at late time,  and imposing that the four-point function is zero at $t=0$, we get, 
 \be \label{J1234}
 \la \da_k \da_{p_2} a_{p_3} a_{p_4} \ra(t) =  4\frac{e^{ i \Delta \omega t} -1}{\Delta \omega} \lam^{*}_{k p_2 p_3 p_4}
 \({1\over n_k} {+} {1\over n_{p_2}} {-} {1\over n_{p_3}} {-} {1\over n_{p_4}}  \) 
n_k n_{p_2} n_{p_3}n_{p_4} ~, \ \ \  \Delta \omega \equiv \omega_k+ \omega_{p_2} - \omega_{p_3}- \omega_{p_4} ~.
 \ee
Substituting (\ref{J1234}) into (\ref{Qkinetic_eq}) yields,
 \be \label{313}
 \frac{\d n_k}{\d t} =  16\Im \sum_{p_2,p_3,p_4} \delta(k{+}p_2{ -} p_3 {-} p_4) |\lam_{k p_2 p_3 p_4}|^2  \frac{e^{i \Delta \omega t}-1}{\Delta \omega} \( {1\over n_k} {+} {1\over n_{p_2}} {-} {1\over n_{p_3}} {-}  {1\over n_{p_4}}  \) 
n_k n_{p_2} n_{p_3}n_{p_4}  ~.
 \ee
 Next, we take the late time limit,  
 \be \label{314}
 \Im \frac{e^{i \Delta \omega t} - 1}{\Delta \omega} = \frac{\sin \Delta \omega t}{\Delta \omega} \rightarrow \pi \delta(\Delta \omega)~ \ \ \ \ \text{as }   \ \ \ \ t\rightarrow \infty
 \ee
 Thus, the equation takes the form,
  \be \label{Q315}
 \frac{\d n_k}{\d t} = 16\pi \sum_{p_2,p_3,p_4} \delta(\Delta \omega)\delta(k{+}p_2 {-} p_3 {-} p_4) |\lam_{k p_2 p_3 p_4}|^2 \( {1\over n_k} { +} {1\over n_{p_2}}  {-} {1\over n_{p_3}} {-}  {1\over n_{p_4}}\) 
n_k n_{p_2} n_{p_3}n_{p_4} ~.
 \ee
 This is the wave kinetic equation \cite{Falkovich}. 
 
 The wave kinetic equation may look familiar; it is reminiscent of the Boltzmann equation. In fact, one can easily redo the above derivation for quantum mechanics, yielding a similar equation but with some of the $n_p$ replaced with $N_p+1$, where $N_p$ now denotes the occupation number, 
  \be \nn
 \frac{\d N_k}{\d t} = 16\pi \sum_{p_2,p_3,p_4} \delta(\Delta \omega)\delta(k{+}p_2 {-} p_3 {-} p_4) |\lam_{k p_2 p_3 p_4}|^2 \( N_{p_3} N_{p_4}(N_k{+}N_{p_2}{+}1) - N_k N_{p_2} (N_{p_3}{+}N_{p_4}{+}1)\) ~.
 \ee
  For large occupation number, $N_k\gg1$, the classical kinetic equation (\ref{Q315}) is recovered, while for small occupation numbers, $N_k\ll 1$, waves behave like particles and one recovers the classical Boltzmann equation. 
 
Returning to the classical kinetic equation (\ref{Q315}),  there is an equilibrium solution,
 \be \label{Qnke}
 n_k = \frac{T}{\omega_k}~.
 \ee
 To see this, notice that with this $n_k$ the terms on the right hand side of  (\ref{Q315})  vanish because they are proportional to $\delta(\Delta \omega)\, \Delta\omega  =0$. The distribution (\ref{Qnke}) is the Rayleigh-Jeans distribution, which is the low energy (high temperature; i.e. classical) limit of the Plank distribution (Bose-Einstein).

\sss*{Turbulent cascade}
A crucial aspect of wave turbulence is that, in addition to the thermal solution (\ref{Qnke}), there is another stationary solution to the kinetic equation (an $n_k$ for which $d n_k/dt=0$). This is the Kolmogorov-Zakharov (KZ) solution.  Since the  kinetic equation   (\ref{Q315})   is only valid at weak nonlinearity, these solutions are called weakly turbulent states \cite{Falkovich}. The KZ solution deviates strongly from equilibrium. Its form depends on the structure of the interaction term and the dispersion relation. For instance, consider homogeneous functions of momenta,
\be
 \omega_p \propto p^\alpha~, \quad \lam_{p_1 p_2 p_3 p_4} =  (p_1 p_2 p_3 p_4)^{\beta\over 4} U \delta_{p_1,p_2; p_3,p_4}~,
 \label{scale_inv_ansatz}
\ee
where $U$ depends only on the ratio of the momenta and their mutual angles. One may check that there is a stationary solution to the kinetic equation, 
\be
 n_p\propto p^{-\gamma}~, \quad \gamma={2\over 3} \beta + d - {\al \over 3} \quad \text{or} \quad \gamma={2\over 3} \beta + d~.
 \label{quarticKZ}
\ee
The conventional way of checking this is through  a change of variables, known as Zakharov's transformation \cite{Zakharov, Falkovich}. A faster method \cite{Polyakov, Gurarie} is the following:  substitute the ansatz $n_p\propto p^{-\gamma}$ into the kinetic equation \reef{Q315} and rescale the momenta $|p_i| = x_i |k|$ for $i=2,3,4$, where $x_i$ are dimensionless parameters, to get, 
\be
  {\partial n_k\over \partial t}   \propto k^{2d-\al-3\gamma+2\beta}~.
\ee 
The proportionality constant is independent of $k$ and its orientation. Comparing with the rotationally symmetric continuity equation in momentum space,\footnote{The continuity equation for the occupation number follows from the $U(1)$ symmetry of the wave system \reef{H21}, $a_k\to e^{i\phi} a_k~, \quad \da_k\to e^{-i\phi} \da_k$.}
\be
 0=  {\partial n_k\over \partial t}    +\vec \nabla\cdot \vec J =  {\partial n_k\over \partial t}   + {1\over k^{d-1}} {\partial\over \partial k} \( k^{d-1} J_k\) ~,
\ee 
 yields
 \be
 k^{d-1} J_k \propto k^{3d - \al-3\gamma+2\beta}~.
 \ee
In particular, the occupation number $n_k$ (or energy spectrum $\omega_k n_k$) is time independent if $k^{d-1} J_k = \text{const}$ (or $k^{d-1} \omega_k J_k = \text{const}$). This condition is equivalent to \reef{quarticKZ}.

\sec{Propagator identities} \label{ap:prop}
In this appendix we collect a few useful propagator identities. The propagator, given in the main body in (\ref{prop38}),  is $D_k(t_{21}) = n_k e^{i \omega_k t_{21} - \gamma_k |t_{21}|}$. Differentiating the propagator gives 
\be
\frac{d}{d t_1}  D_k(t_{21}) = D_k(t_{21}) (- i\omega_k - \gamma_k \sgn(t_{12}))~, \ \ \ \ \ \sgn(t_{12}) = \theta(t_{12}) - \theta(t_{21})~,
\ee
so that, 
\be \label{136}
\sd_{k}(t_1) D_k(t_{21})= F_k e^{-( i\omega_{k}- \gamma_{k})(t_1 - t_2)} \theta(t_{21}) = 2\gamma_k D_k(t_{21}) \theta(t_{21}) ~,
\ee
where we added the time argument $t_1$ to $\sd_k$ to stress that the operator is applied at coordinate $t_1$. Dividing by $F_k$ this is, 
  \be \label{411}
\frac{\sd_k(t_1)}{F_k} D_k(t_{21})= \frac{1}{n_k} D_k(t_{21}) \theta(t_{21}) ~,
\ee
Taking the complex conjugate and using $D_k^*(t_{21})=D_k(t_{12})$, gives
\be \label{412}
 \frac{\bsd_{k}(t_1)}{F_k} D_k(t_{12})=  \frac{1}{n_k} D_k(t_{12}) \theta(t_{21}) ~.
\ee
Finally, an identity that we will need is the action on the propagator of $\bsd_{k}(t_1)$ followed by  $\sd_k(t_2)$. This gives a delta function, 
\be  \label{413}
 \sd_k(t_2) \bsd_k(t_1)  D_k(t_{12})=   F_k \delta(t_{21}) ~.
\ee

\sec{Six-point function} \label{sec:43}
\begin{figure}
  \centering
  \subfloat[]{
    \includegraphics[width=2.6cm]{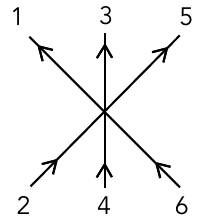}} \ \ \ \ \ \ \ 
      \subfloat[]{
    \includegraphics[width=4cm]{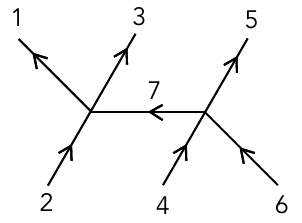}}
       \caption{Tree level diagrams representing the six-point function (\ref{eq:six}).}
        \label{fig:6p}
\end{figure}
In this appendix we give the tree-level six-point function, 
\be \label{eq:six}
\la a_{p_1, \o_1} \da_{p_2, \o_2} a_{p_3, \o_3} \da_{p_4, \o_4}  a_{p_5, \o_5}   \da_{p_6, \o_6} \ra~.
\ee
There are two diagrams: one from the sextic vertex, and one from two quartic vertices, see Fig.\ref{fig:6p}.  Each term by itself is divergent in the limit of zero forcing and zero dissipation, but the sum is finite. To simplify notation we define,
\be \label{Ghatv2}
\hG_i = \frac{1}{G^*_{p_i, \o_i} F_{p_i}}~, \ \ \ i = 1,3,5,7~, \ \ \ \ \ \ \ \hG_i = \frac{-1}{G_{p_i, \o_i} F_{p_i}}~, \ \ \ i =2,4,6~.
\ee
This is slightly different from the convention we used earlier in the one-loop four-point function (\ref{Ghat}). 
The diagram coming from the sextic vertex gives,  
\be \label{sexT}
- 2\pi \delta(\o_{1,3,5; 2,4,6})\, 4\sum_{p_7} \frac{1}{F_{p_7}}\lam_{p_1p_3p_2p_7} \lam_{p_7p_5p_4p_6}  \prod_{i=1}^6 D_{p_i, \o_i}~.
\ee
The diagram coming from two quartic vertices gives, 
\be \label{quartT}
 - 2\pi \delta(\o_{1,3,5; 2,4,6})\, 4
\sum_{p_7} \lam_{p_1p_3p_2p_7} \lam_{p_7p_5p_4p_6} \( \sum_{i=1}^3\hG_i-\hG_7^*  \) 
 \(\sum_{j=4}^6 \hG_j + \hG_7 \) \prod_{i=1}^7 D_{p_i, \o_i} \Big|_{\o_7=\o_5 {+}\o_6{-}\o_{4}}
  \ee
If we expand each of the two parenthesis, there are a total of $16$ terms. The one that involves $\hG_7 \hG_7^*$ is precisely the sextic diagram (\ref{sexT}), with a relative minus sign. Upon adding the two diagrams, (\ref{sexT}) and (\ref{quartT}), these two terms  cancel. 
The sum is therefore, 
 \be \label{426}
 -2\pi \delta(\o_{1,3,5; 2,4,6})\, 4\sum_{p_7} \lam_{p_1p_3p_2p_7} \lam_{p_7p_5p_4p_6} \( \sum_{i=1}^3 \hat G_i \sum_{j=4}^6 \hat G_j -\hat G_7^*\sum_{j=4}^6 \hat G_j +\hat G_7\sum_{i=1}^3 \hat G_i \)\prod_{i=1}^6 D_{p_i, \o_i}~.
 \ee
  The final result for the six-point function is thus (\ref{426}), and permutations thereof: the $3!$ permutations of $(1,3,5)$ and, independently, the $3!$ permutations of $(2,4,6)$. 

\sec{Integrals for the one loop four-point function} \label{ap:integral}
In this appendix we provide some of the details of the computations of the integrals appearing in the one-loop four point function in Sec.~\ref{sec52}. 

\sss*{Frequency space four-point function}
To evaluate (\ref{55}) we open the parenthesis and consider each of the $16$ terms. We need the following integrals, which are fairly trivial contour integrals, 
\be \nn
\!\!\!\! \int\frac{d\o_5}{2\pi} D_{p_5, \o_5} D_{p_6,  \o_3 {+}\o_4{-} \o_5}=  \frac{1}{\o_{3,4;p_5,p_6} {-} i \gamma_{5} {+} i\gamma_{6}}\( \frac{F_6 n_5}{\o_{3,4;p_5,p_6} {-}i \gamma_{56} }+ \frac{F_5 n_6}{\o_{3,4;p_5,p_6} {+}i\gamma_{56} }\)=\frac{ 2n_5 n_6 \g_{56}}{\o_{3,4;p_5,p_6}^2 {+} \gamma_{56}^2}
 \ee
    \be \nn
\int\frac{d\o_5}{2\pi} G_{p_5, \o_5} D_{p_6,  \o_3 {+}\o_4{-} \o_5}= \frac{i\, n_6}{\o_{3,4;p_5,p_6}  {+} i \gamma_{56}}~, \  \ \ \ \  \ \ \ \ \ \ \ \ \  \int\frac{d\o_5}{2\pi} D_{p_5, \o_5} G_{p_6,  \o_3 {+}\o_4{-} \o_5}=  \frac{i\, n_5}{ \o_{3,4;p_5,p_6} {+} i \gamma_{56}}
 \ee
 \be
  \int\frac{d\o_5}{2\pi} G_{p_5, \o_5} G^*_{p_6,  \o_3 {+}\o_4{-} \o_5}=0~, \ \ \ \ \  \ \ \ \  \ \ \ \ \ \ 
    \int\frac{d\o_5}{2\pi} D_{p_5, \o_5}=n_5~,
    \ee
  where $D_{p_i, \o_i}$ was given in (\ref{DG}) and recall that we defined  $\g_{i j} \equiv \g_{p_i} + \g_{p_j}$ and $\g_{ i j k l} \equiv  \g_{p_i} + \g_{p_j} + \g_{p_k}+\g_{p_l}$. 
We now apply this to (\ref{55}). Since $D_{p_i, \o_i} = F_{p_i}|G_{p_i, \o_i}|^2$, we have  e.g. $\hG_1 D_{p_1, \o_1} = G_{p_1, \o_1}$, where $\hG_i$ was given in (\ref{Ghat}). We find that (\ref{55})  becomes, 
 \bml   \label{B2}
-2\pi \delta(\o_{1,2;3,4})\, 8\sum_{p_5,p_6} \lam_{p_1 p_2 p_5 p_6}\lam_{p_5 p_6 p_3 p_4} \prod_{i=1}^4 D_{p_i, \o_i}\Big[ - \(\frac{n_6}{F_{p_5}} + \frac{n_5}{F_{p_6}} \) + \frac{i\, (\hG_3 {+} \hG_4) (n_5 {+} n_6)}{ \o_{3,4;p_5,p_6} {-} i \gamma_{56}}\\ + \frac{i\, (\hG_1 {+}\hG_2)(n_5{+} n_6)}{\o_{3,4;p_5,p_6} {+} i \gamma_{56}} +\frac{ 2 (\hG_1{+}\hG_2)(\hG_3{+}\hG_4) n_5 n_6 \g_{56}}{\o_{3,4;p_5,p_6}^2 {+} \gamma_{56}^2} \Big]~.
\end{multline} 
Forming a common denominator, the result (\ref{511v2}) then follows.  

For evaluating (\ref{515}) we likewise need the following integrals,
\be\nn
 \int\frac{d\o_5}{2\pi} D_{p_5, \o_5} D_{p_6,  \o_3 {+}\o_5{-} \o_1} = \frac{1}{\o_{3,p_5; 1,p_6}{+} i \g_{5}{-}i\g_6}\(\frac{F_6 n_5}{\o_{3,p_5; 1,p_6}{+} i \g_{56}}{+}\frac{F_5 n_6}{\o_{3,p_5; 1,p_6}{-} i \g_{56}}\)=\frac{ 2n_5 n_6 \g_{56}}{\o_{3,p_5; 1,p_6}^2 {+} \gamma_{56}^2}
\ee
\be \nn
  \int\frac{d\o_5}{2\pi} G_{p_5, \o_5} D_{p_6,  \o_3 {+}\o_5{-} \o_1}=\frac{i\, n_6 }{\o_{1,p_6; 3,p_5}  {+} i \gamma_{56}}~, \  \ \ \ \ \  \ \ \  \int\frac{d\o_5}{2\pi} D_{p_5, \o_5} G_{p_6,  \o_3 {+}\o_5{-} \o_1}=\frac{-i\, n_5}{\o_{1, p_6; 3, p_5} {-}i \gamma_{56}}
  \ee
  \be
    \int\frac{d\o_5}{2\pi} G_{p_5, \o_5} G_{p_6,  \o_3 {+}\o_5{-} \o_1}=0~.
    \ee
    Using this  to evaluate the integral in (\ref{515}) gives, 
    \bml   \label{B4}
-2\pi \delta(\o_{1,2;3,4})\, 16\sum_{p_5,p_6}  \lam_{p_1 p_6 p_3 p_5}\lam_{p_2 p_5 p_4 p_6}  \prod_{i=1}^4 D_{p_i, \o_i}\Big[ - \(\frac{n_6}{F_{p_5}} + \frac{n_5}{F_{p_6}} \) + \frac{i\, (\hG_1{+}\hG_3)(n_6{-}n_5)}{\o_{1, p_6; 3, p_5} {+}i \gamma_{56}}\\+ \frac{i\, (\hG_2{+}\hG_4)(n_6{-}n_5)}{\o_{1, p_6; 3, p_5} {-}i \gamma_{56}}  +\frac{ 2 (\hG_1{+}\hG_3)(\hG_2{+}\hG_4) n_5 n_6 \g_{56}}{\o_{1, p_6; 3, p_5}^2 {+} \gamma_{56}^2}  \Big]
\end{multline}
Forming a common denominator gives (\ref{figb}). 

\sss*{Equal time four-point function}
In Sec.~\ref{sec52}, after finding the frequency space one-loop four-point function, we found the equal-time one-loop four-point function, see (\ref{4ptloopPos}). Here we record some of the necessary contour integrals to get this answer. The equal-time four-point function is found by Fourier transform of the frequency space four-point function, see (\ref{FT4}). We will be taking the Fourier transform, with respect to $\o_1, \o_2, \o_3, \o_4$, of (\ref{B2}) and (\ref{B4}).  Due to the energy conserving delta function in the frequency-space, $\delta(\o_{1,2; 3,4})$, the exponential in the Fourier transform drops out (i.e. the equal time four-point function is independent of time) and we simply need to take the integral with respect to $\o_1, \o_2, \o_3, \o_4$  of (\ref{B2}) and (\ref{B4}). 

Starting with (\ref{B2}), we need integrals such as, 
\be
\int \prod_{i=1}^4 \frac{d\o_i}{2\pi} 2\pi \delta(\o_{1,2;3,4})\prod_{i=1}^4 D_{p_i, \o_i}\frac{i\, \hG_1}{\o_{3,4;p_5,p_6} {+} i \gamma_{56}}  = n_2 n_3 n_4\frac{i}{\o_{p_3, p_4; p_1, p_2} {+} i \g_{1234}}\frac{i}{\o_{p_3, p_4; p_5, p_6}{+}i \g_{3456}}
\ee
To get this, we first did the $\o_1$ integral by using the delta function. We then did the $\o_2$ integral by closing the contour in the lower half plane and picking up the pole at  $\o_2 = \o_{p_2} - i \gamma_{p_2}$. Likewise, we do the $\o_3$ integral to pick up the pole at  $\o_3 = \o_{p_3} + i \gamma_{p_3}$, and the $\o_4$ integral to pick up the pole at $\o_4 = \o_{p_4} + i \gamma_{p_4}$. One could have of course closed the contours in either the upper or lower half plane; we simply described the easiest choice. By interchange of $1\leftrightarrow 2$,  we therefore have, 
\be  \label{B6}
\int\! \prod_{i=1}^4 \frac{d\o_i}{2\pi} 2\pi \delta(\o_{1,2;3,4})\prod_{i=1}^4\! D_{p_i, \o_i}\frac{i\, (\hG_1+ \hG_2)}{\o_{3,4;p_5,p_6} {+} i \gamma_{56}}  = \prod_{i=1}^4 n_i\!\(\frac{1}{n_1} {+} \frac{1}{n_2}\)\!\frac{i}{\o_{p_3, p_4; p_1, p_2} {+} i \g_{1234}}\frac{i}{\o_{p_3, p_4; p_5, p_6}{+}i \g_{3456}}
\ee
Recall from the definition of $\hG_i$ in (\ref{Ghat}) that $D_{p_1, \o_1} \hG_1 = G_1$ and $D_{p_2, \o_2} \hG_2 = G_2$, while $D_{p_3, \o_3} \hG_3 = -G_3^*$ and $D_{p_4, \o_4} \hG_4 = -G_4^*$. Therefore, by taking (\ref{B6}) and exchanging $1,2 \leftrightarrow 3,4$ and complex conjugating, we get, 
\be  \label{B7}
\int\! \prod_{i=1}^4 \frac{d\o_i}{2\pi} 2\pi \delta(\o_{1,2;3,4})\prod_{i=1}^4\! D_{p_i, \o_i}\frac{i\, (\hG_3+ \hG_4)}{\o_{3,4;p_5,p_6} {-} i \gamma_{56}}  = \prod_{i=1}^4 n_i\!\(\frac{1}{n_3} {+} \frac{1}{n_4}\)\!\frac{i}{\o_{p_1, p_2; p_3, p_4} {-} i \g_{1234}}\frac{i}{\o_{p_1, p_2; p_5, p_6}{-}i \g_{1256}}
\ee
These two integrals, (\ref{B6}) and (\ref{B7}), explain the second line of the four-point function answer (\ref{4ptloopPos}). 

Another integral we need is, 
\bml
\int \prod_{i=1}^4 \frac{d\o_i}{2\pi} \, 2\pi \delta(\o_{1,2;3,4})\prod_{i=1}^4 D_{p_i, \o_i}\frac{  2\hG_1\hG_3 \g_{56}}{\o_{3,4;p_5,p_6}^2 {+} \gamma_{56}^2} \\
= - n_2 n_4 \frac{i}{\o_{p_3,p_4; p_5, p_6}{+}i \g_{3456}} \frac{i}{\o_{p_5,p_6; p_1, p_2} {+}i \g_{1256}}\(1+\frac{2 \g_{56} i}{\o_{p_3,p_4; p_1, p_2}{+}i \g_{1234}} \)~,
\end{multline}
from which, by symmetry, it follows that, 
\bml
\int \prod_{i=1}^4 \frac{d\o_i}{2\pi}\,  2\pi \delta(\o_{1,2;3,4})\prod_{i=1}^4 D_{p_i, \o_i}\frac{  2(\hG_1{+}\hG_2)(\hG_3{+}\hG_4)  \g_{56}}{\o_{3,4;p_5,p_6}^2 {+} \gamma_{56}^2}\\
=  \prod_{i=1}^4 n_i \(\frac{1}{n_1}{+}\frac{1}{n_2} \) \(\frac{1}{n_3} {+} \frac{1}{n_4}\) \frac{-i}{\o_{p_3,p_4; p_5, p_6}{+}i \g_{3456}} \frac{i}{\o_{p_5,p_6; p_1, p_2} {+}i \g_{1256}}\( 1+\frac{2 \g_{56} i}{\o_{p_3,p_4; p_1, p_2}{+}i \g_{1234}} \)
\end{multline}
This explains the first line of the four-point function answer (\ref{4ptloopPos}). 

Now, turning to (\ref{B4}) we need a few integrals:
\be
\int \prod_{i=1}^4 \frac{d\o_i}{2\pi} 2\pi \delta(\o_{1,2;3,4})\prod_{i=1}^4 D_{p_i, \o_i}\frac{i\, \hG_1}{\o_{1, p_6; 3, p_5} {+}i \gamma_{56}} = n_2 n_3 n_4\frac{i}{\o_{p_4, p_6; p_2, p_5} {+}i \gamma_{2456}}\frac{i}{\o_{p_3, p_4; p_1, p_2}{+}i \g_{1234}}~,
\ee
which simply follows by using the delta function to do the  $\o_1$ integral, and then closing  the contours for the $\o_2, \o_3, \o_4$ integrals in such a way so as to only ever pick poles at $\o_i = \o_{p_i} \pm i \g_i$. Another integral we need can be obtained from this one through interchange $1, 2, 5 \leftrightarrow 3,4,6$ combined with complex conjugation.~\footnote{Since $D_{p_1, \o_1} \hG_1 = G_1$ and  $D_{p_3, \o_3} \hG_3 = -G_3^*$, when we exchange $p_1, \o_1$ with $p_3, \o_3$ and complex conjugate, we shouldn't forget about the minus sign.}
\be\nn
\!\int \!\prod_{i=1}^4 \frac{d\o_i}{2\pi} 2\pi \delta(\o_{1,2;3,4})\!\prod_{i=1}^4 D_{p_i, \o_i}\frac{i\, (\hG_1 + \hG_3)}{\o_{1, p_6; 3, p_5} {+}i \gamma_{56}} = \prod_{i=1}^4 n_i\! \(\frac{1}{n_1} {-}\frac{1}{n_3}\) \frac{i}{\o_{p_4, p_6; p_2, p_5} {+}i \gamma_{2456}}\frac{i}{\o_{p_3, p_4; p_1, p_2}{+}i \g_{1234}}
\ee
Exchanging $1,3,5\leftrightarrow 2,4,6$ gives, 
\be \nn
\!\int\! \prod_{i=1}^4 \frac{d\o_i}{2\pi} 2\pi \delta(\o_{1,2;3,4})\!\prod_{i=1}^4 D_{p_i, \o_i}\frac{-i\, (\hG_2 + \hG_4)}{\o_{1, p_6; 3, p_5} {-}i \gamma_{56}} = \prod_{i=1}^4 n_i\! \(\frac{1}{n_2} {-}\frac{1}{n_4}\) \frac{i}{\o_{p_3, p_5; p_1, p_6} {+}i \gamma_{3516}}\frac{i}{\o_{p_3, p_4; p_1, p_2}{+}i \g_{1234}}
\ee
This explains the fourth line of (\ref{4ptloopPos}). Finally we need, 
\bml 
\int \prod_{i=1}^4 \frac{d\o_i}{2\pi} \, 2\pi \delta(\o_{1,2;3,4})\prod_{i=1}^4 D_{p_i, \o_i}\frac{  2\hG_1\hG_2 \g_{56}}{\o_{1, p_6; 3, p_5}^2 {+} \gamma_{56}^2}\\
= n_3 n_4 \frac{i}{\o_{p_4,p_6; p_2, p_5}{+}i \g_{2456}} \frac{i}{\o_{p_3,p_5; p_1, p_6} {+}i \g_{1356}}\(1+\frac{2 \g_{56} i}{\o_{p_3,p_4; p_1, p_2}{+}i \g_{1234}} \)~,
\end{multline}
\bml
\int \prod_{i=1}^4 \frac{d\o_i}{2\pi} \, 2\pi \delta(\o_{1,2;3,4})\prod_{i=1}^4 D_{p_i, \o_i}\frac{  2\hG_1\hG_4 \g_{56}}{\o_{1, p_6; 3, p_5}^2 {+} \gamma_{56}^2}\\
= -n_2 n_3 \frac{i}{\o_{p_4,p_6; p_2, p_5}{+}i \g_{2456}} \frac{i}{\o_{p_3,p_5; p_1, p_6} {+}i \g_{1356}}\(1+\frac{2 \g_{56} i}{\o_{p_3,p_4; p_1, p_2}{+}i \g_{1234}} \)~,
\end{multline}
which, by symmetry, gives us, 
\bml \nn
\int \prod_{i=1}^4 \frac{d\o_i}{2\pi}\,  2\pi \delta(\o_{1,2;3,4})\prod_{i=1}^4 D_{p_i, \o_i}\frac{  2(\hG_1{+}\hG_3)(\hG_2{+}\hG_4) \g_{56}}{\o_{1, p_6; 3, p_5}^2 {+} \gamma_{56}^2}\\
=  \prod_{i=1}^4 n_i \(\frac{1}{n_1}{-}\frac{1}{n_3} \) \(\frac{1}{n_2} {-} \frac{1}{n_4}\) \frac{i}{\o_{p_4,p_6; p_2, p_5}{+}i \g_{2456}} \frac{i}{\o_{p_3,p_5; p_1, p_6} {+}i \g_{1356}}\(1+\frac{2 \g_{56} i}{\o_{p_3,p_4; p_1, p_2}{+}i \g_{1234}} \)~.
\end{multline}
This explains the third line of (\ref{4ptloopPos}).

\section{Higher order kinetic equation details} \label{apE}
In this appendix we fill in some details relevant to simplifying the next-to-leading order kinetic equation.

We start with (\ref{kins}), which we would like to simplify, 
\bml  \label{E1}
  \({\partial n_k(t)\over \partial t}\)_{\text{one-loop}}^a  = -32\sum_{p_1, \ldots, p_6}\delta_{k, p_1}
   \Big[\text{Re}( \lam_{p_3 p_4 p_1 p_2} \lam_{p_1 p_2 p_5 p_6}\lam_{p_5 p_6 p_3 p_4} ) V_{p \delta}(1,2,3,4,5,6) \\ + \text{Im}( \lam_{p_3 p_4 p_1 p_2} \lam_{p_1 p_2 p_5 p_6}\lam_{p_5 p_6 p_3 p_4} ) ( V_{\delta \delta}(1,2,3,4,5,6)+V_{pp}(1,2,3,4,5,6)) \Big]~. 
\end{multline}
Let us start with the first term in (\ref{E1}). We write out $V_{p \delta}$ in terms of $A_{p\delta}$, as defined in (\ref{429}), 
\bea \nn
&&\!\!\!\!\!\!\!\!\!\!\!\!\sum_{p_1, \ldots, p_6}\!\delta_{k, p_1}\text{Re}( \lam_{p_3 p_4 p_1 p_2} \lam_{p_1 p_2 p_5 p_6}\lam_{p_5 p_6 p_3 p_4} ) V_{p \delta}(1,2,3,4,5,6) \\
&=&\!\! \!\!\!\!\sum_{p_1, \ldots, p_6}\!\delta_{k, p_1}\text{Re}( \lam_{p_3 p_4 p_1 p_2} \lam_{p_1 p_2 p_5 p_6}\lam_{p_5 p_6 p_3 p_4} )\! \( A_{p \delta}(1,2,3,4,5,6){ -}  A_{p \delta}(3,4,5,6,1,2)  {-}  A_{p \delta}(5,6,1,2,3,4) \) \nn\\
&=&\!\! \!\!\!\! \sum_{p_1, \ldots, p_6}\! \delta_{k, p_1}\text{Re}( \lam_{p_3 p_4 p_1 p_2} \lam_{p_1 p_2 p_5 p_6}\lam_{p_5 p_6 p_3 p_4} ) 2A_{p \delta}(1,2,3,4,5,6) ~, \label{E2}
  \eea
  where to get to the second line we used that
  \bea \nn
  &&\sum_{p_1, \ldots, p_6}\delta_{k, p_1}\text{Re}( \lam_{p_3 p_4 p_1 p_2} \lam_{p_1 p_2 p_5 p_6}\lam_{p_5 p_6 p_3 p_4} )A_{p \delta}(3,4,5,6,1,2) \\ \nn
  &=& \sum_{p_1, \ldots, p_6}\delta_{k, p_1}\text{Re}( \lam_{p_3 p_4 p_1 p_2} \lam_{p_1 p_2 p_5 p_6}\lam_{p_5 p_6 p_3 p_4} )(-A_{p \delta}(5,6,3,4,6,1,2) ) \\
 & =& -\sum_{p_1, \ldots, p_6}\delta_{k, p_1}\text{Re}( \lam_{p_3 p_4 p_1 p_2} \lam_{p_1 p_2 p_5 p_6}\lam_{p_5 p_6 p_3 p_4} )A_{p \delta}(3,4,5,6,1,2) = 0~,
  \eea
  where to get the first equality we use that $ A_{p \delta}(3,4,5,6,1,2) $  is antisymmetric under $(3,4)\leftrightarrow (5,6)$, and then to get the second we do a change of dummy variables $(3,4)\leftrightarrow (5,6)$ and use that the real part of the product of couplings is symmetric under this change. 
  We also used that for the last term on the second line of (\ref{E2}) we have, 
  \bea \nn
   &&\sum_{p_1, \ldots, p_6}\delta_{k, p_1}\text{Re}( \lam_{p_3 p_4 p_1 p_2} \lam_{p_1 p_2 p_5 p_6}\lam_{p_5 p_6 p_3 p_4} ) (- A_{p \delta}(5,6,1,2,3,4)) \\ \nn
   & =&   \sum_{p_1, \ldots, p_6}\delta_{k, p_1}\text{Re}( \lam_{p_3 p_4 p_1 p_2} \lam_{p_1 p_2 p_5 p_6}\lam_{p_5 p_6 p_3 p_4} )  A_{p \delta}(1,2,5,6,3,4)  \\
   &=&\sum_{p_1, \ldots, p_6}\delta_{k, p_1}\text{Re}( \lam_{p_3 p_4 p_1 p_2} \lam_{p_1 p_2 p_5 p_6}\lam_{p_5 p_6 p_3 p_4} )  A_{p \delta}(1,2,3,4,5,6)~,
   \eea
   where to get the first equality we used the antisymmetry of $A_{p, \delta}$, and to get the second we did a change of dummy variables $(3,4)\leftrightarrow (5,6)$. This now explains the result in (\ref{E2}). 

Now we turn to the second term in (\ref{E1}). We have that, 
\bea \nn
&&\!\!\!\!\!\!\!\!\!\!\!\!\sum_{p_1, \ldots, p_6}\!\delta_{k, p_1}\text{Im}( \lam_{p_3 p_4 p_1 p_2} \lam_{p_1 p_2 p_5 p_6}\lam_{p_5 p_6 p_3 p_4} ) V_{\delta \delta}(1,2,3,4,5,6) \\ \nn
&=&\!\!\!\!\!\! \sum_{p_1, \ldots, p_6}\!\delta_{k, p_1}\text{Im}( \lam_{p_3 p_4 p_1 p_2} \lam_{p_1 p_2 p_5 p_6}\lam_{p_5 p_6 p_3 p_4} ) \(A_{\delta\delta}(1, 2,3,4,5,6){ -}  A_{\delta\delta}(3,4,5,6,1,2) {-}  A_{\delta\delta}(5,6,1,2,3,4)\)\\ 
&=& \!\!\!\!\!\!\sum_{p_1, \ldots, p_6}\!\delta_{k,p_1} \text{Im}( \lam_{p_3 p_4 p_1 p_2} \lam_{p_1 p_2 p_5 p_6}\lam_{p_5 p_6 p_3 p_4} )2 A_{\delta \delta}(1,2,3,4,5,6)~,
\eea
where the $A_{\delta\delta}(3,4,5,6,1,2)$ contribution vanished because $A_{\delta\delta}(3,4,5,6,1,2)$ is symmetric under $(3,4)\leftrightarrow (5,6)$ whereas the imaginary part of the product of couplings is antisymmetric, and the $A_{\delta\delta}(5,6,1,2,3,4)$ terms is equal to the $A_{\delta\delta}(1, 2,3,4,5,6) $ term by writing $A_{\delta\delta}(5,6,1,2,3,4) = A_{\delta\delta}(1,2,5,6,3,4)$ and then doing a change of variables $(3,4)\leftrightarrow (5,6)$. 

 Finally, for the third term in (\ref{E1}) we have that
\bea \nn
&&\!\!\!\!\!\!\!\!\!\!\!\!\sum_{p_1, \ldots, p_6}\!\delta_{k, p_1}\text{Im}( \lam_{p_3 p_4 p_1 p_2} \lam_{p_1 p_2 p_5 p_6}\lam_{p_5 p_6 p_3 p_4} ) V_{p p}(1,2,3,4,5,6) \\ \nn
&=&\!\!\!\!\!\!  \sum_{p_1, \ldots, p_6}\!\delta_{k, p_1}\text{Im}( \lam_{p_3 p_4 p_1 p_2} \lam_{p_1 p_2 p_5 p_6}\lam_{p_5 p_6 p_3 p_4} )\(A_{pp}(1, 2,3,4,5,6) +  A_{pp}(3,4,5,6,1,2) +  A_{pp}(5,6,1,2,3,4)\)\\
&= &0~,
\eea
because the $A_{pp}(3,4,5,6,1,2) $ contribution vanishes since  $A_{pp}(3,4,5,6,1,2) $ is symmetric under interchange $(3,4)\leftrightarrow (5,6)$ whereas the imaginary part of the product of couplings is antisymmetric, while for $A_{pp}(5,6,1,2,3,4)$ term we use that $A_{pp}(5,6,1,2,3,4) = A_{pp}(1,2,5,6,3,4)$ and then do a change of variables $(3,4)\leftrightarrow (5,6)$. As a result of antisymmetry of the  imaginary part of the product of couplings, this term the cancels the $A_{pp}(1, 2,3,4,5,6) $ term. 

So, we finally have that (\ref{E1}) is 
\bml
  \({\partial n_k(t)\over \partial t}\)_{\text{one-loop}}^a  = -64\sum_{p_1, \ldots, p_6}\delta_{k, p_1} 
   \Big[\text{Re}( \lam_{p_3 p_4 p_1 p_2} \lam_{p_1 p_2 p_5 p_6}\lam_{p_5 p_6 p_3 p_4} ) A_{p \delta}(1,2,3,4,5,6) \\ + \text{Im}( \lam_{p_3 p_4 p_1 p_2} \lam_{p_1 p_2 p_5 p_6}\lam_{p_5 p_6 p_3 p_4} )A_{\delta \delta}(1,2,3,4,5,6) \Big]~.
\end{multline}

{\setstretch{1}

\bibliographystyle{utphys}

\end{document}